\documentclass[traditabstract]{aa}

\usepackage{graphicx}
\usepackage{txfonts}
\usepackage{natbib}
\usepackage{aas_macros}
\usepackage{lscape}
\usepackage{longtable}

\usepackage{soul}

\usepackage[usenames,dvipsnames]{color}

\begin{document}

\title{Single Stellar Populations in the Near-Infrared\\ I. Preparation of the IRTF spectral stellar library}
\titlerunning{Single Stellar Populations in the Near-IR - I. Preparation of the IRTF spectral stellar library}

\author{S. Meneses-Goytia\inst{1} 
\and 
R. F. Peletier\inst{1} 
\and 
S. C. Trager\inst{1} 
\and 
J. Falc{\'o}n-Barroso\inst{2,3} 
\and 
M. Koleva\inst{4} 
\and
A. Vazdekis\inst{2,3} 
}
\authorrunning{S. Meneses-Goytia et al.}

\institute{Kapteyn Instituut, Rijksuniversiteit Groningen, Landleven 12, 9747AD Groningen, The Netherlands \\
\email{s.meneses-goytia@astro.rug.nl}
\and
Instituto de Astrof{\'i}sica de Canarias, via L{\'a}ctea s/n, La Laguna, Tenerife, Spain
\and
Departamento de Astrof{\'i}sica, Universidad de La Laguna, 38205 La Laguna, Tenerife, Spain
\and
Sterrenkundig Observatorium, Ghent University, Krijgslaan 281, S9, 9000 Ghent, Belgium
}

\date{Accepted XXXX. Received XXXX; in original form XXXX}

\abstract{
We present a detailed study of the stars of the {\it IRTF spectral library} to understand its full extent and reliability for use with Stellar Population (SP) modeling. The library consist of 210 stars, with a total of 292 spectra, covering the wavelength range of $0.94$ to $2.41~\mu m$ at a resolution $R \approx 2000$. For every star we infer the effective temperature ($T_{\mathrm{eff}}$), gravity ($\log g$) and metallicity ($[Z/Z_{\odot}]$) using a full-spectrum fitting approach in a section of the K band ($2.19$ to $2.34~\mu m$) and temperature-NIR colour relations. We test the flux calibration of these stars by calculating their integrated colours and comparing them with the Pickles library colour-temperature relations. We also investigate the NIR colours as a function of the calculated effective temperature and compared them in colour-colour diagrams with the Pickles library. This latter test shows a good broad-band flux calibration, important for the SP models. Finally, we measure the resolution $R$ as a function of wavelength. We find that the resolution increases as a function of lambda from about $6~\AA$ in $J$ to $10~\AA$ in the red part of the $K$-band.  With these tests we establish that the IRTF library, the largest currently available general library of stars at intermediate resolution in the NIR, is an excellent candidate to be used in stellar population models. We present these models in the next paper of this series.
}


\maketitle


\label{firstpage}

\section{Introduction}
\hspace{0.45cm}

The near-infrared (NIR) spectral region contains several features that provide information about the stellar content of galaxies. The spectra of stellar populations in the NIR are strongly influenced by cool, late-type stars. Red giant branch (RGB) stars are old ($> 2~\mathrm{Gyr}$) stars which have a stronger contribution at older ages and also as a function of redder wavelengths. Thermally pulsating asymptotic giant branch (AGB) stars, on the other hand, contribute most to the integrated light of a stellar population between $1$ and $3~\mathrm{Gyr}$. However, during the life-time of a galaxy, we also find regular AGB stars contributing to the spectrum at all ages. Therefore, in order to create stellar population models for all galaxy ages,  we need stellar libraries that include these stars. 

Stellar libraries are compilations of stellar spectra with a certain wavelength range and resolution covering a large part of the parameter space of effective temperature, gravity and chemical abundances. These libraries can be either theoretical or empirical. Theoretical libraries are calculated using models and atmospheres that can be determined for a wide range of stellar parameters and detailed chemical abundances, at nearly unlimited spectral resolution \citep[e.g.][among others]{westera_et_al_2002,coelho_et_al_2007,allard_et_al_2012}. Model stellar atmospheres, however, are at present not able to reproduce the full spectra of some observed stars as a result of systematic uncertainties in our understanding of stellar atmospheres and a lack of complete atomic and molecular line lists, especially at lower temperatures. Observational or empirical libraries are compilations of observations of real stars that come with instrumental limitations such as limited resolution, incomplete correction of atmospheric absorption, and since most of the spectra come from stars in the Galactic disk, they might not be fully adequate to study stellar populations in other environments, such as elliptical galaxies. Examples of often used stellar libraries include \citet{lancon_and_wood_2000}, \citet{cenarro_et_al_2001c}, and \citet{sanchez-blazquez_et_al_2006a}.

Since hot stars dominate the light in the Ultraviolet (UV), while cool stars are prominent in the Near-Infrared (NIR), it is advantageous to use a large wavelength range when studying the stellar populations of galaxies \citep{frogel_et_al_1978,maraston_2005}. Until now, most empirical stellar libraries are found in the optical, while some are available in the near-UV, and only very few in the NIR. With these libraries one has been able to model stellar populations in globular clusters and galaxies in the optical in two ways, using line indices, mostly using the Lick-IDS system \citep[e.g.][]{burstein_et_al_1984,worthey_et_al_1994}, and using full-spectrum fitting. Excellent full-spectrum fits have been made of dwarf galaxies \citep{koleva_et_al_2009}, elliptical galaxies \citep{yamada_et_al_2006,conroy_and_van_dokkum_2012}, and globular clusters \citep{schiavon_et_al_2004}. The fits of giant elliptical galaxies show that some strong lines, most prominently the Mg $b$ line at $5177~\AA$,  cannot be fit by stellar population models using empirical libraries, indicating that the [Mg/Fe] ratios in these objects are different from that of the solar neighborhood \citep{peterson_1977,peletier_1989,worthey_et_al_1992b}. By now it is clear that many abundance ratios in galaxies are not the same as in the solar neighborhood (e.g. Yamada et al. 2006, Conroy \& van Dokkum 2013). To first order galaxy spectra in the optical between $4000$ and $6000~\AA$ are, however, fairly well understood. 

\begin{figure}[!tb]
	\centering
	\includegraphics[width=\columnwidth]{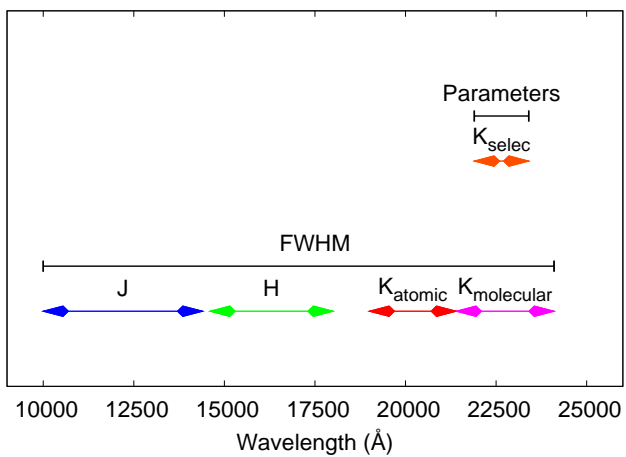}
	\caption{Wavelength ranges used in this work. For the determination of the stellar parameters through full-spectrum fitting, we use a section of the $K$ band ($2.19$ to $2.34~\mu m$, see Section \,\ref{irtf_fsf}). For the determination of the nominal spectral resolution (Section \,\ref{fwhm}), we divide the wavelength range into four sections representing respectively the $J$ ($1.04-1.44~\mu m$) and $H$ ($1.46-1.80~\mu m$) bands, as well as the {\it atomic} ($1.90-2.14~\mu m$) and {\it molecular}$-$dominated ($2.14-2.41~\mu m$) ranges of the $K$ band.}
	\label{lambda_ranges}
\end{figure}

Theoretical model atmospheres have become more and more detailed over the years, producing libraries in different wavelength ranges. The most complete libraries in the NIR include the BaSeL models \citep{lejeune_et_al_1997,lejeune_et_al_1998,westera_et_al_2002}, Tlusty models \citep{lanz_and_hubeny_2003,lanz_and_hubeny_2007}, and the models by \citet{aringer_et_al_2009}. Compendia of stellar-type spectra formed from spectra of stars of similar type have also been made, for example by \citet{pickles_1998}, that combine various observations providing standard spectra for all spectral types and luminosity classes. Empirical libraries in the NIR include \citet{lancon_and_wood_2000}, \citet{cushing_et_al_2005} and \citet{rayner_et_al_2009}. \citet{mouhcine_and_lancon_2002}, \citet{maraston_et_al_2009b} and \citet{conroy_and_gunn_2009} have made stellar population models based on these libraries. 

Stellar spectra observed by \citet{rayner_et_al_2009} and \citet{cushing_et_al_2005}, compiled in the {\it IRTF spectral library} allow the modelling of stellar populations in the NIR. \citet{conroy_and_gunn_2009} used a handful of spectra of this library and \citet{meneses_and_peletier_2012} have constructed preliminary SSP models with this library. The library is a powerful ingredient for stellar population models in the NIR due to the large number of cool stars it contains. These stars contribute strongly in this wavelength range and their spectra are of higher resolution than their empirical predecessors in the same wavelength region such as those observed by  \citet{lancon_and_wood_2000}. To be able to use the {\it IRTF spectral library} for stellar population synthesis, an accurate determination of the spectral calibration and associated stellar parameters of the stars in the library must be established. A well-calibrated stellar library, with an accurate set of stellar parameters and known resolution, leads to a reliable development of models. The information that can be extracted from these models, such as integrated colours are linked to the colours of the stars.

\begin{figure}[!htb]
	\centering
	\includegraphics[width=1.0\columnwidth]{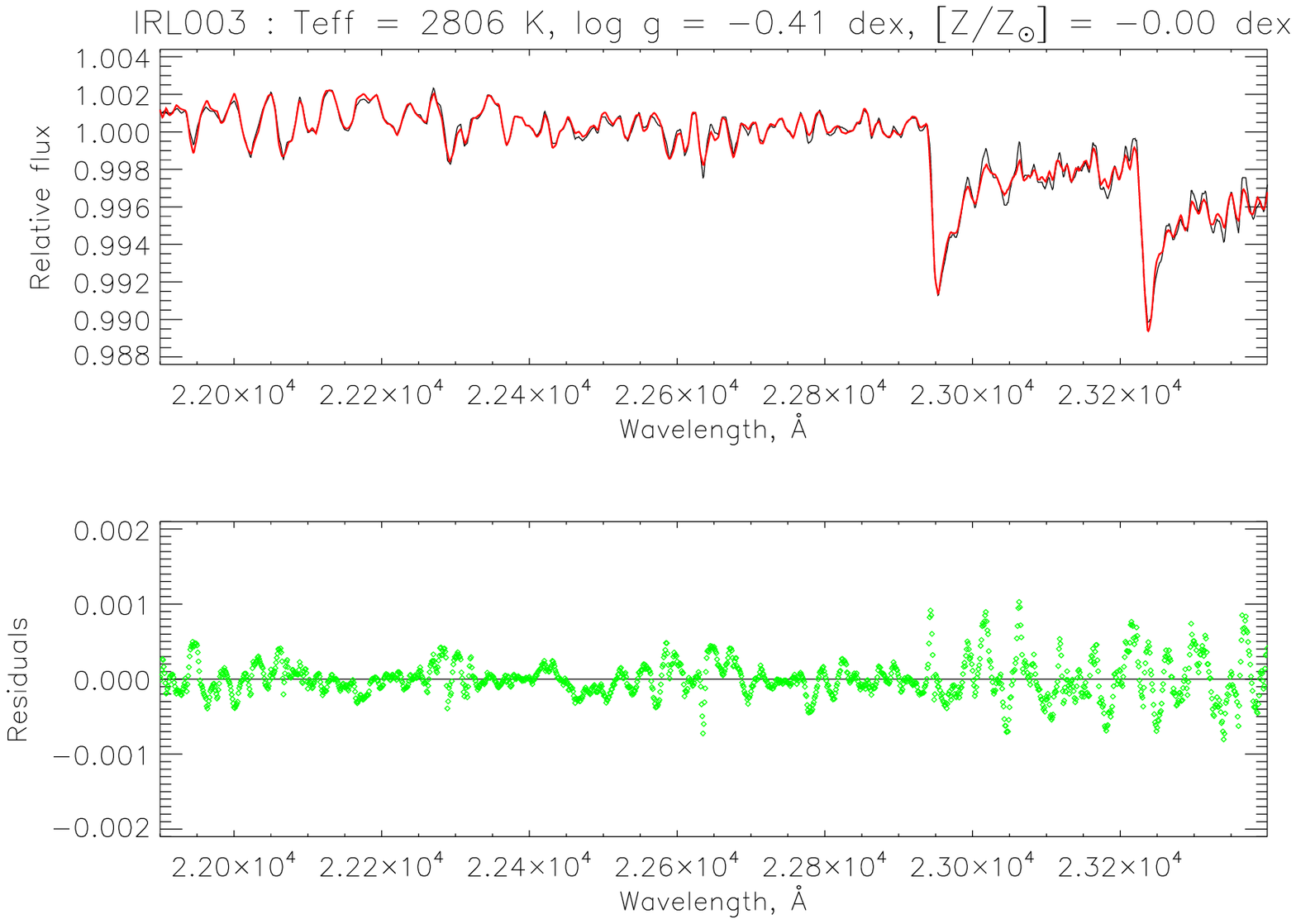}
	\includegraphics[width=1.0\columnwidth]{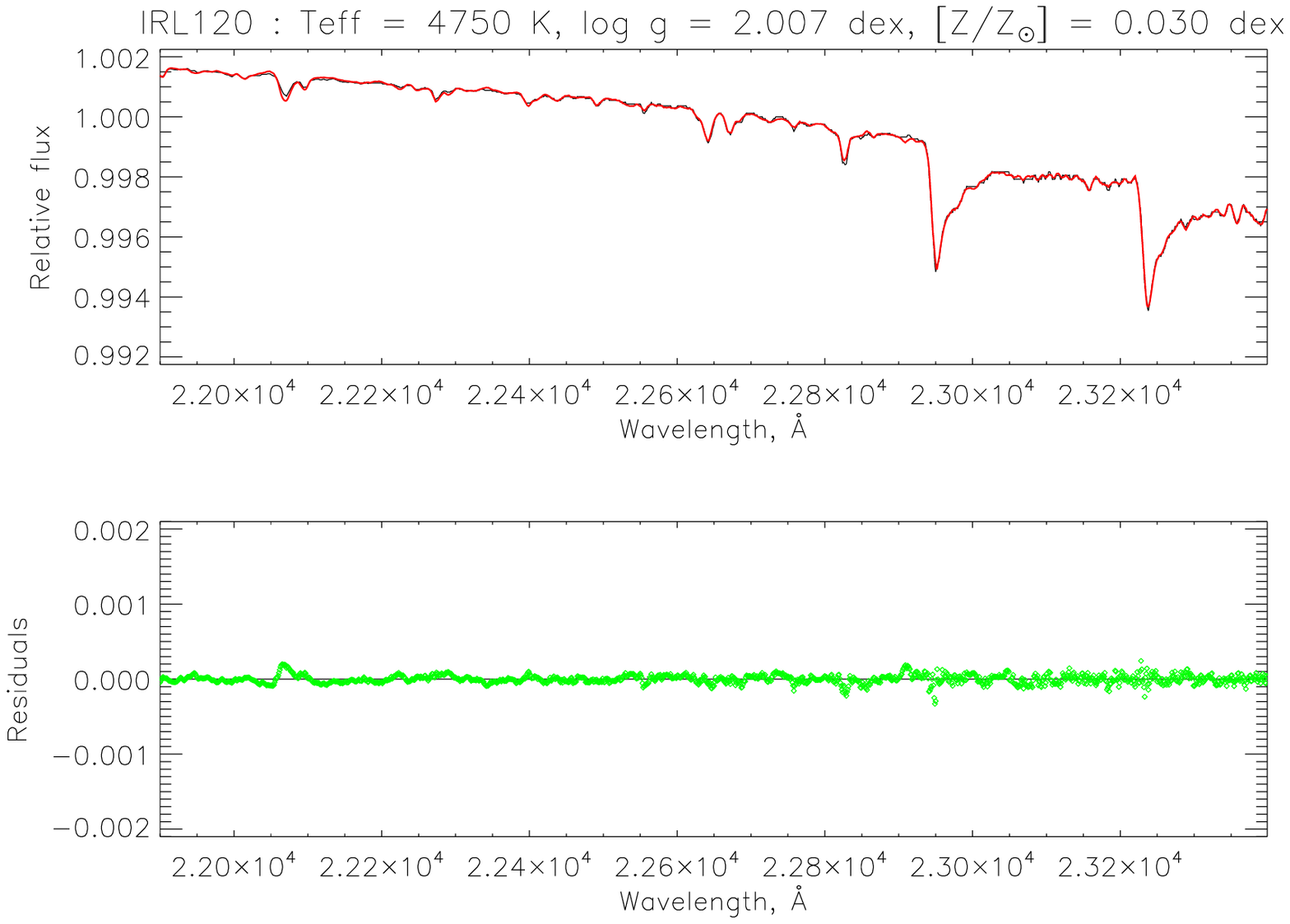}
	\includegraphics[width=1.0\columnwidth]{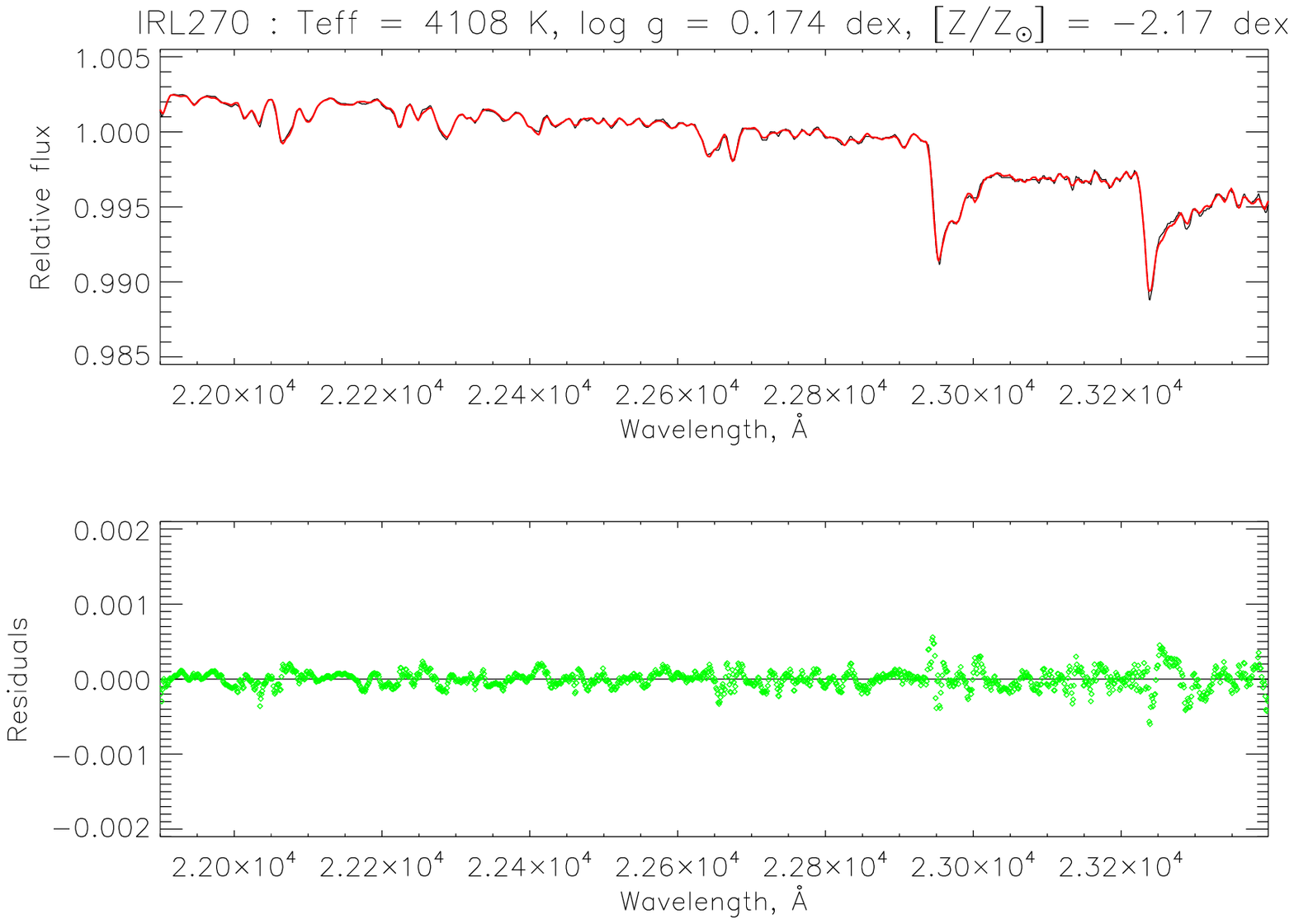}
	\caption{Example of the results obtained with full-spectrum fitting. The name of the stars and the obtained parameters are cited in figure titles. In the upper panels, we show the observed spectrum (black lines) and the best fit model (red lines).  In the lower panels, we show the residuals (green data points) which for these stars are less than 0.2$\%$. For details of the fits, see text.}
	\label{irtf_fitting}
\end{figure}

In this series of papers, we aim to understand more about the cool stellar populations in unresolved galaxies by building Single Stellar Population (SSP) models in the NIR and comparing them to galaxy observations. In this paper, the first of this series, we characterise the spectra of the {\it IRTF spectral library} and determined the stellar parameters, $T_{\mathrm{eff}}$, $\log g$ and metallicity ($[Z/Z_{\odot}]$) of the stars, tied to the parameters in \citet{cenarro_et_al_2007}, in Section \,\ref{parameters}. We test the flux calibration of the library and the behaviour of the integrated colours of its stars and their stellar parameters in Section \,\ref{irtf_tests}. And finally, in Section \,\ref{fwhm}, we measure the Full Width at Half Maximum (FWHM) and resolution of the library spectra. In Figure \,\ref{lambda_ranges}, we show the wavelength ranges used for our analysis. 

\begin{figure}[!htb]
	\centering
	\includegraphics[width=1.\columnwidth]{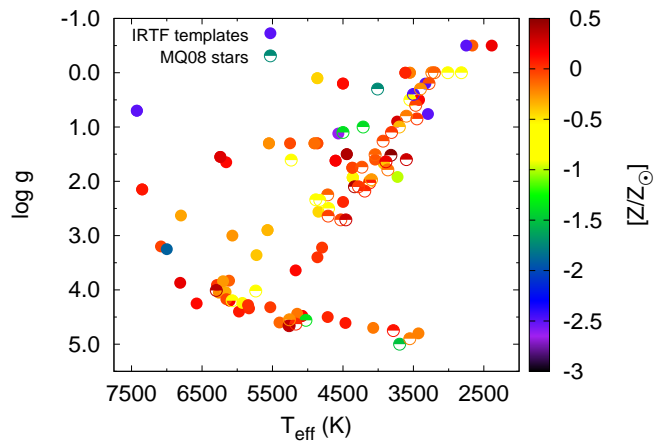}
	\caption{HR diagram of the stars that compose the template library for the determination of stellar parameters of the {\it IRTF spectral library}. This template set is formed by $73$ IRTF stars in common with the MILES and CaT empirical stellar libraries, $52$ additional stars observed by \citet[][MQ08]{marmol-queralto_et_al_2008}.}
	\label{irtf_HR}
\end{figure}

In our second paper \citep[][hereafter Paper II]{paper_II}, we calculate SSP models following a similar approach to \citet{vazdekis_et_al_2010} but using the {\it IRTF spectral library}. We calculate full Spectral Energy Distributions (SEDs), by finding a stellar spectrum with the appropriate stellar parameters ($T_{\mathrm{eff}}$, $\log g$ and metallicity) from the stellar library using interpolation for every point on a theoretical isochrone, weighting them by a initial mass function. In the third paper of the series \citep[][hereafter Paper III]{paper_III}, we use our models to analyse a number of composite stellar systems.

\begin{figure}[!htb]
	\centering
	\includegraphics[width=1.1\columnwidth]{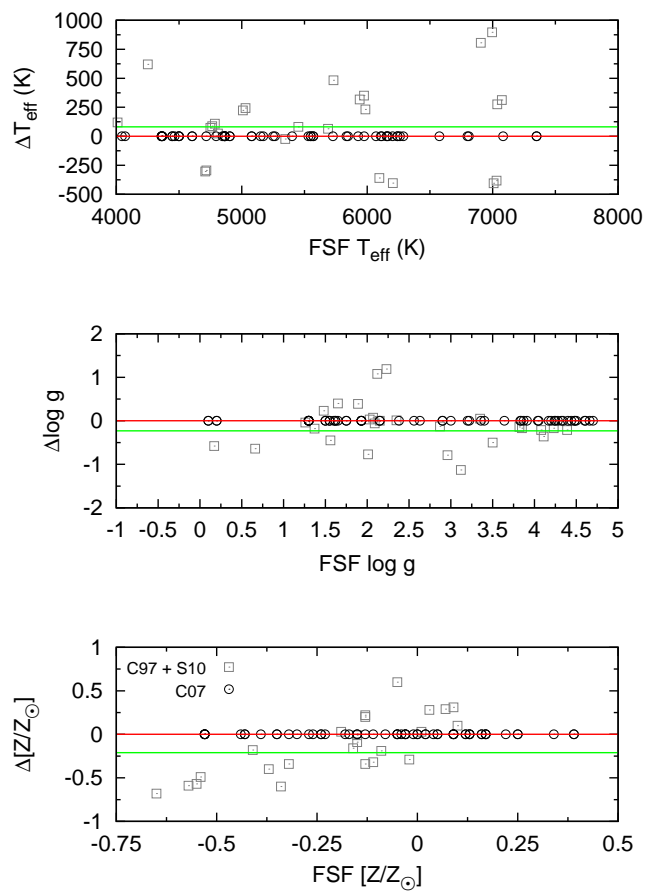}
	\caption{Residuals from the comparison of the stellar parameters obtained from the full-spectrum fitting (FSF) approach and the literature values from C07, S10 and C97 and the residuals (FSF - Lit). The red line is the one-to-one correlation. The green line marks the median of the difference 81K for $T_{\mathrm{eff}}$, -0.23 for $\log g$ and -0.07 of metallicity) which represents the offset between non-C07 and C97 + S10 literature values.}
	\label{lit_comparisons}
\end{figure}


\section{The IRTF spectral library}
\hspace{0.45cm}

The {\it IRTF spectral library\footnote{\url{irtfweb.ifa.hawaii.edu/~spex/IRTF_Spectral_Library/}}} is a compilation of stellar spectra observed with the medium-resolution spectrograph SpeX at the NASA Infrared Telescope Facility on Mauna Kea \citep{rayner_et_al_2009, cushing_et_al_2005}. This library covers the NIR range from $0.8$ to $2.5~\mu m$~(and extends in some cases out to $5.2~\mu m$). We focus on the spectral region for the $J$, $H$ and $K$ bands ($0.94$ to $2.41~\mu m$). These spectra were observed at a resolving power of $R$ $=$ 2000 ($R = \lambda/\Delta\lambda$, see below), and their continua were not normalized, keeping the strong molecular absorption features from cool stars. Keeping the spectral shape also allowed relative flux calibration between the stars, by scaling the spectra to published Two Micron All Sky Survey (2MASS) photometry ($J$, $H$ and $K_{s}$ magnitudes), implying that the integrated colours obtained from the spectra are consistent with those obtained by 2MASS. 

\begin{figure}[!htb]
	\centering
	\includegraphics[width=1.\columnwidth]{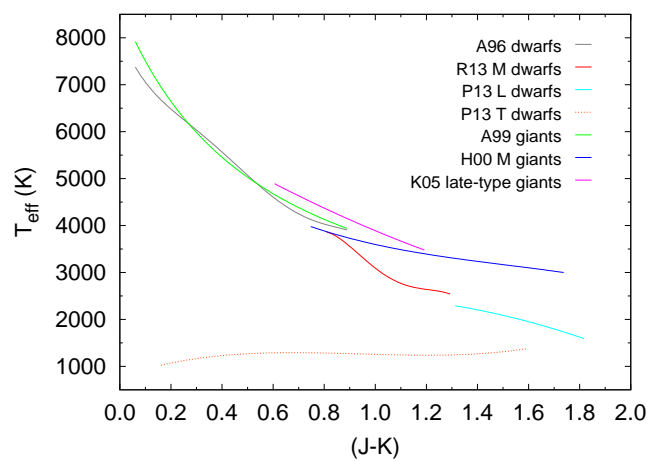}
	\caption{Colour-temperature relations for different regimes of effective temperature and luminosity class from \citet[][A96]{alonso_et_al_1996}, \citet[][R13]{rajpurohit_et_al_2013}, \citet[][P13]{pecaut_and_mamajek_2013}, \citet[][A99]{alonso_et_al_1999}, \citet[][H00]{houdashelt_et_al_2000} and \citet[][K05]{kucinskas_et_al_2005}.}
	\label{authors_HR_CTR}
\end{figure}
 
\begin{figure}[!htb]
	\centering
	\includegraphics[width=\columnwidth]{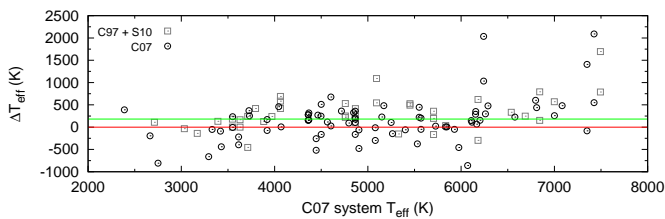}
	\caption{Residuals from the comparison of the stellar parameters obtained from the colour-temperature relation (CTR) approach and the literature values from C07, S10 and C97 and the residual (Lit $_{C10~system}$ - CTR). The red line is the one-to-one correlation. The green line marks the median of the difference (181K) which represents the offset in temperatures between the  CTR method and literature in the C07 system.}
	\label{FSF_CTR}
\end{figure}

The spectral types of the 210 library stars include F, G, K, M, L, S and C types, with luminosity classes from  supergiants (I) to dwarfs (V). The majority of the stars are cool stars, which dominate the light in the NIR. Around $60\%$ of the stars are variable, including Cepheids, RR Lyrae and semi-regular variables. Additionally, some stars have two spectra, corrected and non-corrected for extinction. We kept these spectra as individual stars leaving us with a sample of 292 spectra.


\section{Determination of stellar parameters}
\label{parameters}
\hspace{0.45cm}

A crucial step towards using a spectral stellar library for SSP modelling (see Paper II) is to accurately know the atmospheric parameters of its stars to be able to tie them to the evolutionary tracks during the modelling. Moreover, the parametric coverage of the library is vital to understand the applicable ranges in the models constructed from its stars. For this purpose, we apply two methods, one in which the $T_{\mathrm{eff}}$, $\log g$ and metallicity of many of the stars of the {\it IRTF spectral library} are determined in a self-consistent way from a sample of stars with well-determined parameters (see Section \,\ref{irtf_fsf}) and another in which we use the colour$-$temperature relations for different regimes (see Section \,\ref{irtf_temp_color}). We use parameters from the literature to establish the validity of these methods.

\begin{figure}[!htb]
	\centering
	\includegraphics[width=1.\columnwidth]{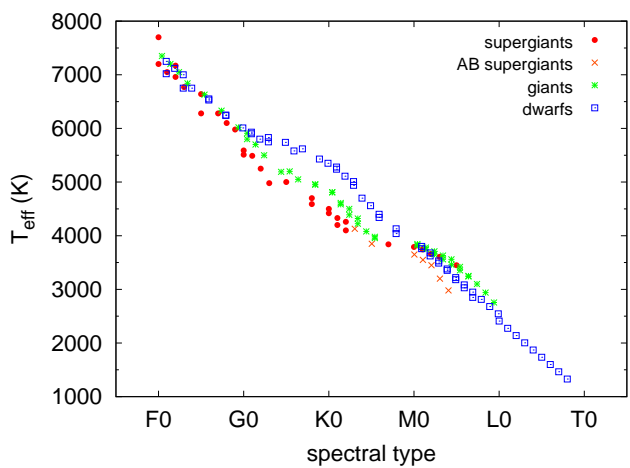}
	\caption{Compilation of effective temperatures as a function of spectral type from \citet{gray_and_corbally_2009} and \citet{cox_1999}.}
	\label{irtf_spec}
\end{figure}


\subsection{Full$-$spectrum fitting method}
\label{irtf_fsf}
\hspace{0.45cm}
Here we use the $\it{ULySS}\footnote{\url{ulyss.univ-lyon1.fr}}$ package \citep{koleva_et_al_2009} to compare the IRTF spectra to a template library with known parameters, and to find the set of best-matching spectra. Briefly, {\it{ULySS}} fits a spectrum with a linear combination of non-linear components (in this case stellar spectra) convolved with a line-of-sight velocity distribution and multiplied by a polynomial continuum. 

The parameters of the best-matching spectra along with their respective weights give the atmospheric parameters of the IRTF stars. Figure \,\ref{irtf_fitting} shows examples of the fits for three stars (IRL003, IRL120 and IRL270) with their corresponding resulting parameters. It is important to mention that the full$-$spectrum fitting (FSF) approach used for measuring the atmospheric parameters was only used on a small region of the $K$ band ($2.19$ to $2.34~\mu$m). 

We choose as our primary template library 73 stars of the {\it IRTF spectral library} that are also found in the empirical stellar libraries MILES \citep{sanchez-blazquez_et_al_2006a} and CaT \citep{cenarro_et_al_2001c}. The parameters of those stars are determined using a compilation from the literature by \citet[][hereafter C07]{cenarro_et_al_2007}. Additionally, we use as templates 52 stars observed in a section of the $K$ band ($2.19$ to $2.34~\mu$m) by \citet{marmol-queralto_et_al_2008} which are also contained in the MILES and CaT libraries and for which there are also stellar atmospheric parameters available in C07.

\begin{figure}[!htb]
	\centering
	\includegraphics[width=1.\columnwidth]{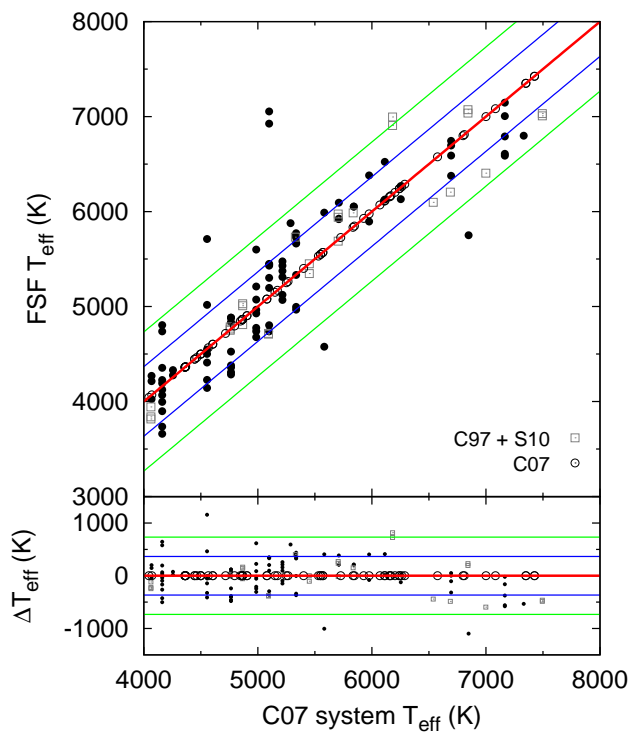}
	\caption{Comparison of the effective temperature obtained from the full-spectrum fitting (FSF) with literature values in the C07 system. The squares and circles are those stars with literature values from either C07, S10 or C97 already in the C07 system, for which parameters are computed using their stellar type. The black data points are stars that do not have literature values in C07, S10 or C97. The red line is the one-to-one relation. The blue and green lines mark the $1~\sigma$ and $2~\sigma$ confidence intervals, respectively. The residuals (FSF - Lit $_{C07~system}$) are presented in the lower panel.}
	\label{irtf_Teff_selection}
\end{figure}

\begin{figure}[!t]
	\centering
	\includegraphics[width=1.\columnwidth]{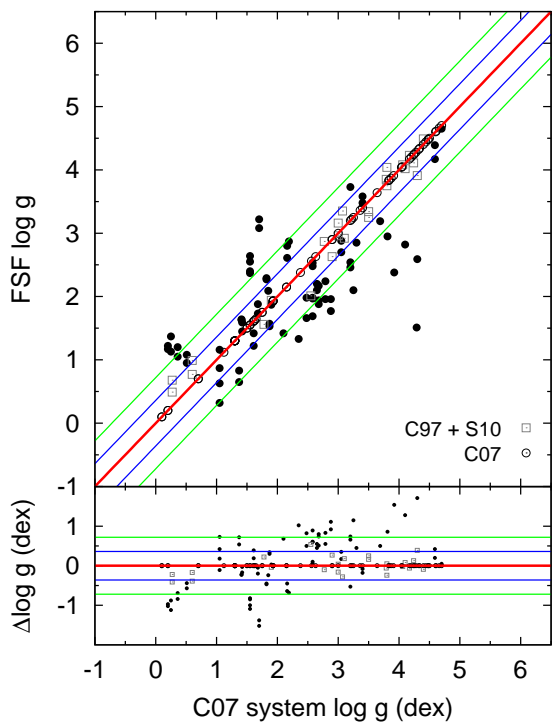}
	\caption{As in Figure \,\ref{irtf_Teff_selection} but for $\log g$.}
	\label{irtf_logG_selection}
\end{figure}

\begin{figure}[!b]
	\centering
	\includegraphics[width=1.\columnwidth]{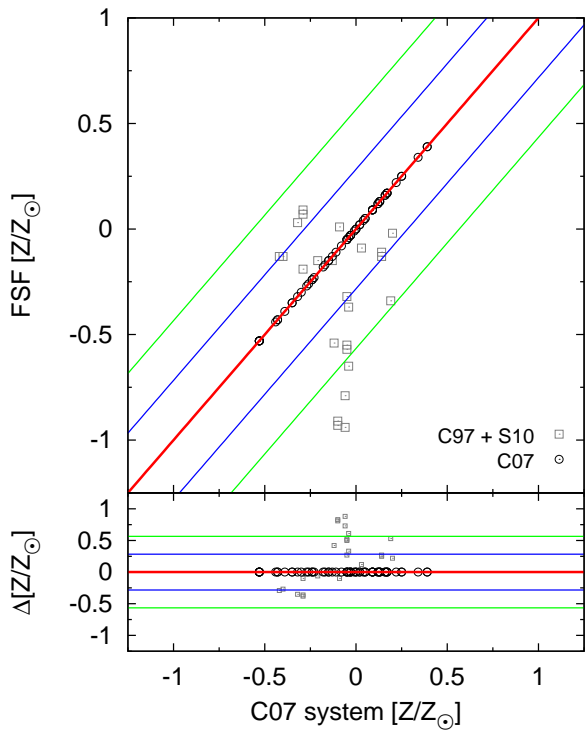}
	\caption{As in Figure\,\ref{irtf_Teff_selection} but for the stars with literature values ($[Z/Z_{\odot}]$). Note that the diagram contains fewer stars than e.g. Figure\,\ref{irtf_logG_selection}. This is because there are many stars without literature values, for which there is no way to determine their matallicity. For these stars we have assumed solar chemical abundances and not plotted them here.}
	\label{irtf_Metal_selection}
\end{figure}

For our analyses, we also create 108 interpolated stars whose parameters are selected from grids of different $T_{\mathrm{eff}}$, $\log g$ and 4 metallicities that were within the limits provided by the empirical stars of our template library. Their temperature ranges from $2500$ to $6500\,\mathrm{K}$, with steps of $500\,\mathrm{K}$ and their gravities range from $-0.25$ to $4.0\,\mathrm{dex}$, with steps of $1.0\,\mathrm{dex}$. The metallicities considered here are $-0.70$, $-0.4$, $0.0$ and $0.2\,\mathrm{dex}$. The spectrum of each interpolated star is obtained by interpolating between empirical spectra of stars with known parameters in our template library. We apply an interpolation scheme similar to that of \citet{vazdekis_et_al_2003}, which creates a box of stellar parameters around a given point with certain stellar parameters, i.e. the parameters of the star for which we want to obtain a spectrum. This box is flexible enough to expand until a sufficient number of adequate stars are found, if necessary, and is divided into eight cubes of different sizes that have the given point as a corner. This minimises the errors due to the lack of stars in certain regions of the distribution of stars. The size of the box is inversely proportional to the density of stars found around the point and the box can be as small as the typical uncertainties of the parameters of the template library. When the boxes are determined, the spectra of the stars that form each one of the eight boxes are combined into 8 different spectra. As a final step, these spectra are interpolated to obtain a final  spectrum with the desired stellar parameters. This step is taken since the available interpolators with $\it{ULySS}$ are only for optical wavelengths. We subsequently use the $\it{ULySS}$ package, along with the 233 template stars described above to determine stellar atmospheric parameters in the full-spectrum approach.

\begin{figure}[t]
	\centering
	\includegraphics[width=1.\columnwidth]{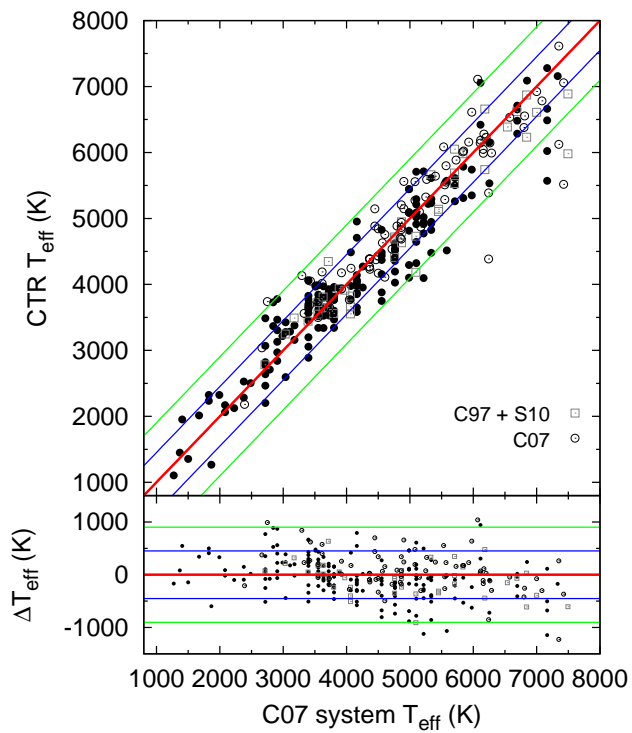}
	\caption{As in Figure \,\ref{irtf_Teff_selection} but here showing the effective temperature obtained from the $(J-K) - T_{\mathrm{eff}}$ relations shown in Figure \,\ref{authors_HR_CTR}.}
	\label{irtf_Teff_selection_color}
\end{figure}

We have run the FSF for the 73 IRTF-MILES stars (hereby C07-stars) to not only test this method but also to homogenise our "anchor" literature values. We have also applied the methodology to the IRTF stars with literature values from \citet[][hereafter S10]{soubiran_et_al_2010} and \citet[][hereafter C97]{cayrel_de_strobel_et_al_1997}, giving a combined sample of 40 stars hereby designated non-C07 stars. 

The FSF method is not applied to stars with $T_{\mathrm{eff}}$ below 4000$\,\mathrm{K}$ due to the lack of templates covering an adequate grid of atmospheric parameters for cool stars. Therefore, below 4000$\,\mathrm{K}$ we adopt atmospheric parameters from the literature or the colour-temperature relation method (see Section\,\ref{irtf_temp_color}).

Figure\,\ref{lit_comparisons} shows that the parameters obtained for the non-C07 stars differ from the catalogs; therefore we calculate the offset between these values from the median of the difference 81K for $T_{\mathrm{eff}}$, -0.23 for $\log g$ and -0.07 for metallicity). We apply the corresponding offset for each parameter to the literature values for the non-C07 stars in order to homogenise the "anchor" stellar atmospheric parameters in the C07 system.

From our analyses we see that most of the stars are in the metallicity range between $\approx$$-$0.7 and 0.2, indicating that the {\it IRTF spectral library} is useful for NIR SP modelling mostly when working with populations near solar abundance.


\begin{figure}[!t]
	\centering
	\includegraphics[width=\columnwidth]{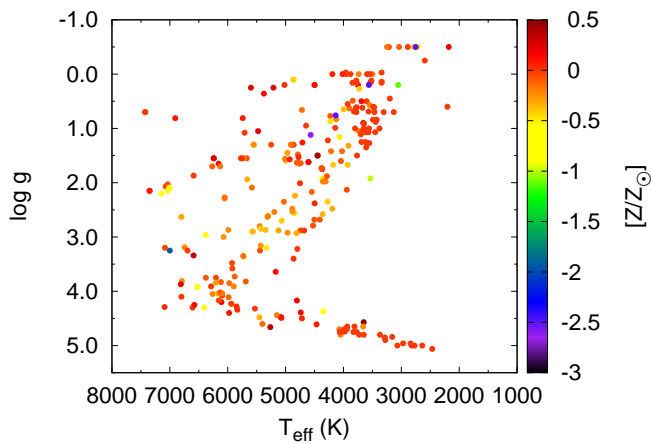}
	\caption{The stellar atmospheric parameters for the stars of the {\it IRTF spectral library}. This shows the parameter coverage of this library for stellar population models. L and T-type dwarfs were not included.}
	\label{irtf_HR_final}
\end{figure}

\begin{figure}[!b]
	\centering
	\includegraphics[width=\columnwidth]{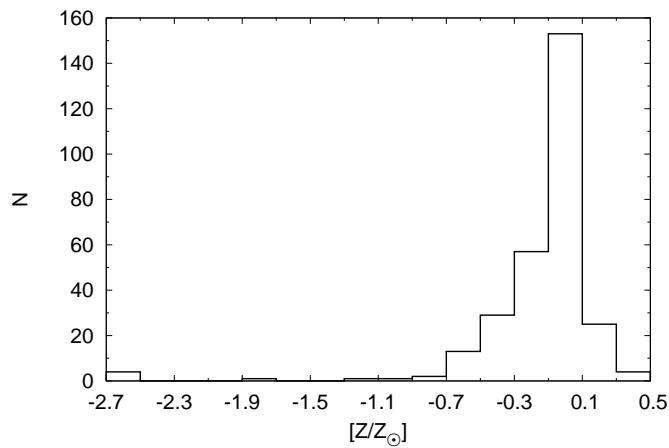}
	\caption{The metallicity distribution function for the stars of the {\it IRTF spectral library}.}
	\label{irtf_mdf}
\end{figure}

\subsection{$T_{\mathrm{eff}}$ and NIR$-$colour relations}
\label{irtf_temp_color}
\hspace{0.45cm}
For this method, we determine the temperature as a function of NIR colours. For this purpose, we use colour-temperature relations (CTR) in the NIR, specifically $(J-K)$, for different regimes of effective temperatures and luminosity classes. For giants, we use the relations of \citet[][A99]{alonso_et_al_1999}, \citet[][H00]{houdashelt_et_al_2000} and \citet[][K05]{kucinskas_et_al_2005} which are applied to F0-K5, M0-M7 and M8-cooler stars, respectively. For dwarfs, the relations of \citet[][A96]{alonso_et_al_1996}, \citet[][R13]{rajpurohit_et_al_2013} and \citet[][P13]{pecaut_and_mamajek_2013} are used for F0-K5, M0-M10 and L-T stars, respectively. We show these relations and their behaviour in Figure\,\ref{authors_HR_CTR}. 

Before applying these relations to those stars without literature parameters, we use them to re-calculate the values of the C07 and non-C07 stars in order to determine the difference between the literature values in the C07 system and colour-temperature relation results and anchor the latter values to the former. This comparison and the respective median offset are shown in Figure\,\ref{FSF_CTR}. This offset~ (181K) is applied to the effective temperatures from the CTR approach. Note that large difference of literature and CTR temperatures of very bright giants is due to the variable colours as a result of stellar pulsations.


\begin{figure}[!b]
	\centering
	\includegraphics[width=\columnwidth]{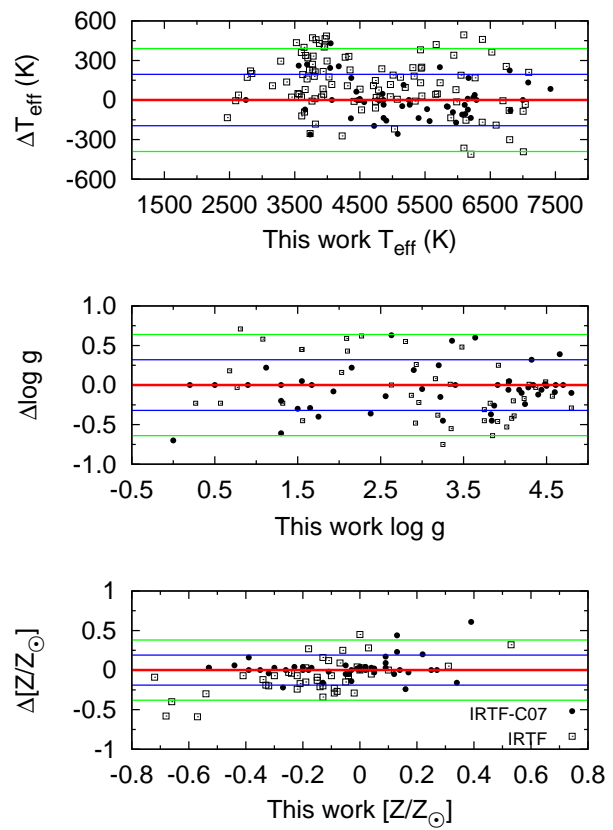}
	\caption{Residuals from the comparison between the stellar atmospheric parameters for the stars of the {\it IRTF spectral library} and those compiled by \citet{cesetti_et_al_2013} (Present work - Cesetti). The circles represent the values for the IRTF stars with C07 stellar parameters and the squares the IRTF stars for which we determined their parameters. The red line is the one-to-one correlation. The blue and green lines mark the $1~\sigma$ and $2~\sigma$ confidence intervals, respectively.}
	\label{irtf_cesetti}
\end{figure}

\subsection{Selection of the atmospheric parameters}
\label{irtf_final_param}
\hspace{0.45cm}
To assess the accuracy of the parameters we obtained with both methods, we compare the parameters of the remaining 219 IRTF stars with literature values from the catalogues of S10 and C97. When literature data are not available, we infer the stellar temperature of the stars from their stellar classification by compiling information from \citet{gray_and_corbally_2009} and \citet{cox_1999}, as shown in Figure \,\ref{irtf_spec}. To determine surface gravity, we use evolutionary tracks based on empirical stars as a function of temperature, gravity and spectral type from \citet{straizys_and_kuriliene_1981}. In Table \,\ref{nomenclature}, we show our calculated parameters, including those from the literature, and those of the template stars.

The agreement between the measured parameters by the FSF method and the literature is reasonable (as shown in Figures \,\ref{irtf_Teff_selection}, \,\ref{irtf_logG_selection} and \,\ref{irtf_Metal_selection}), with $\sigma_{T_{eff}} = 366\,\mathrm{K}$, $\sigma_{\log g}= 0.36\,\mathrm{dex}$ and $\sigma_{[Z/Z_{\odot}]}= 0.28\,\mathrm{dex}$. This technique was limited by the parameter coverage of the template library which was dominated by solar metallicity giants and dwarfs between $\sim 3500$ and $7500~K$. The stars for which we could not determine good parameters are late-type giants (M6 and cooler) and M, L and T dwarfs ($T_{eff} < 2500\,\mathrm{K}$). 

In Figure \,\ref{irtf_Teff_selection_color} we show the resulting effective temperatures from the colour-temperature relations and the comparison with the literature values. This method also shows a good agreement with $\sigma_{T_{eff}} =451\,\mathrm{K}$, except for a few very bright giants whose colours oscillate due to stellar pulsations.

\begin{figure}[!t]
	\centering
	\includegraphics[width=0.9\columnwidth]{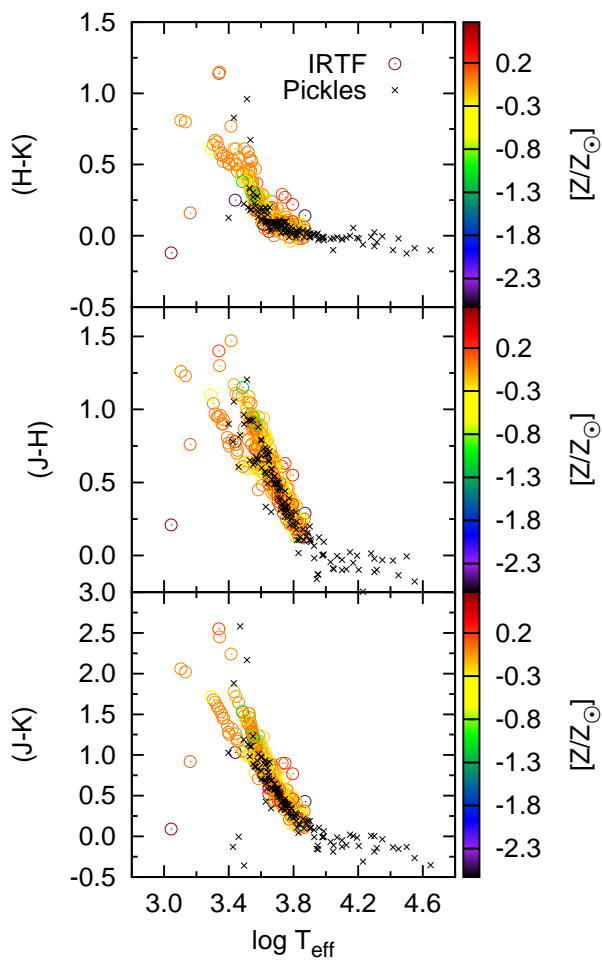}
	\caption{Behaviour of the integrated colours of the {\it IRTF spectral library} stars as a function of effective temperature, compared with the Pickles stellar library (which is assumed to have solar metallicity).}
	\label{irtf_teff_colors}
\end{figure}

It is important to point out that these methods are independent from each other. The FSF determines the parameters of a given star by comparing its spectrum to an empirical spectral library whose stellar parameters were compiled from the literature \citep[for details of this compilation see][]{cenarro_et_al_2007}. On the other hand, the CTR is based on the relation of observed photometry of stars and their effective temperature \citep[see e.g.][]{alonso_et_al_1999}.

To compute the Stellar Population models (Paper II), we mainly use the parameters mainly from the full-spectrum fitting and for those stars whose values are not in the $2~\sigma_{T_{eff}}$ confidence interval, we take the results from the colour-temperature relation that are also within the corresponding reliable limits. For $\log g$, we take the FSF results, or, if they are not within the $2~\sigma_{\log g}$ limits, we take the literature values. We prefer not to use the mass-luminosity relations to calculate the $\log g$ because it requires assigning a particular age and metallicity making the method subject to a bias. For outliers in metallicity, literature values were used if available, otherwise solar metallicity was chosen.

\begin{figure}[!t]
	\centering
	\includegraphics[width=\columnwidth]{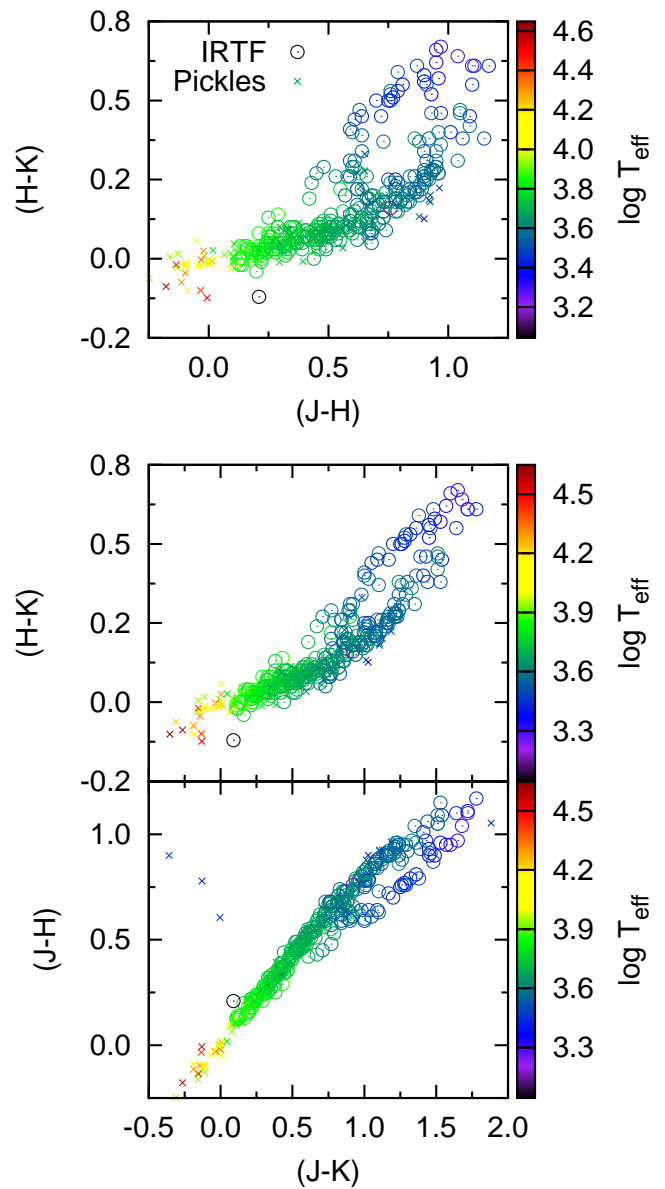}
	\caption{Colour-Colour diagrams of the stars of the {\it IRTF spectral library}, compared with the Pickles library (which is assumed to have solar metallicity).}
	\label{irtf_color_color}
\end{figure}

The calculated values within the $2~\sigma$ confidence intervals (for the respective parameter) are shown in Table \,\ref{nomenclature}. Figure \,\ref{irtf_HR_final} shows the HR diagram of these stars and Figure \,\ref{irtf_mdf} their metallicity distribution function.


\begin{figure*}[!htb]
	\centering
	\includegraphics[width=0.925\columnwidth]{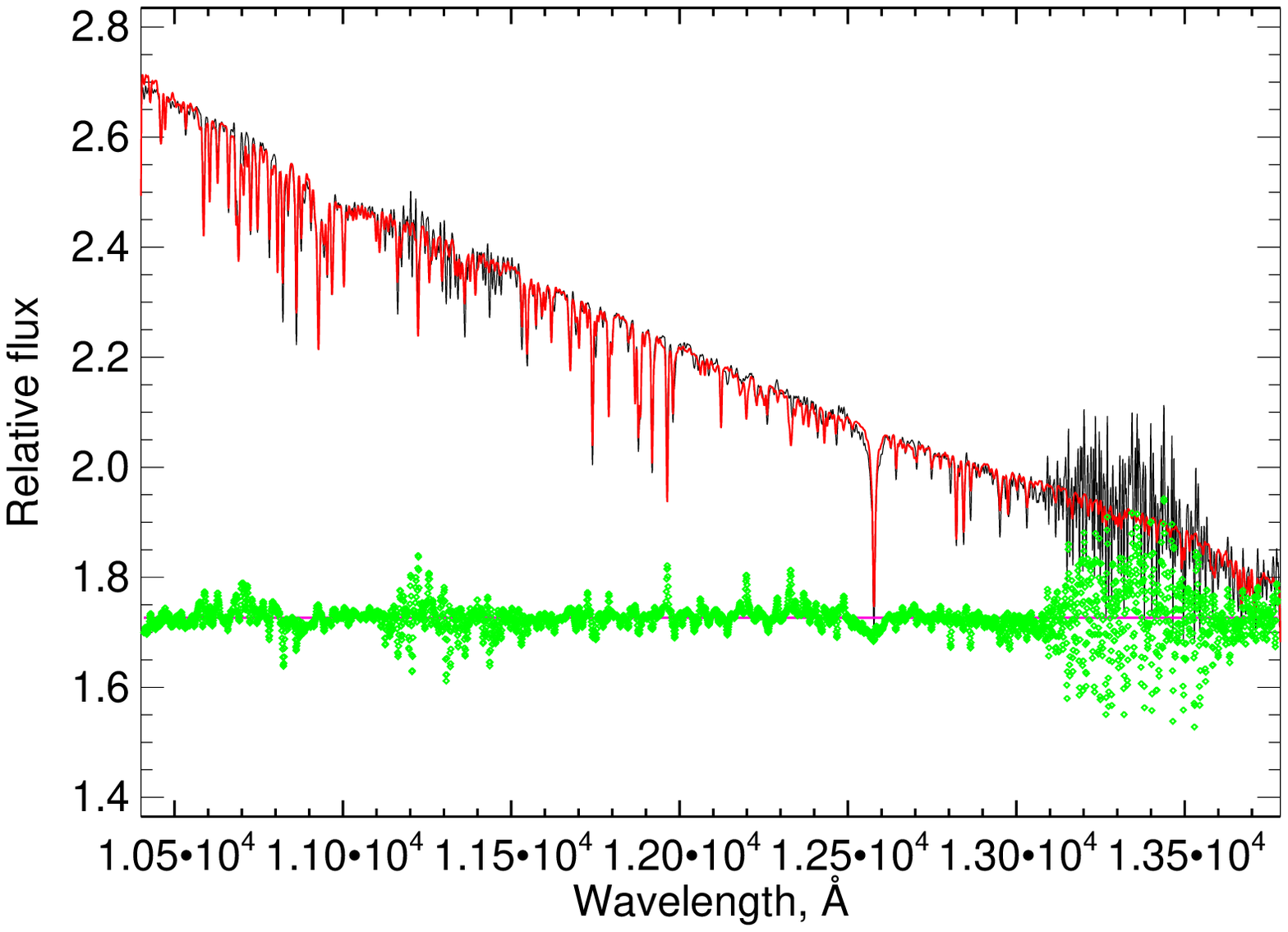}
	\includegraphics[width=0.925\columnwidth]{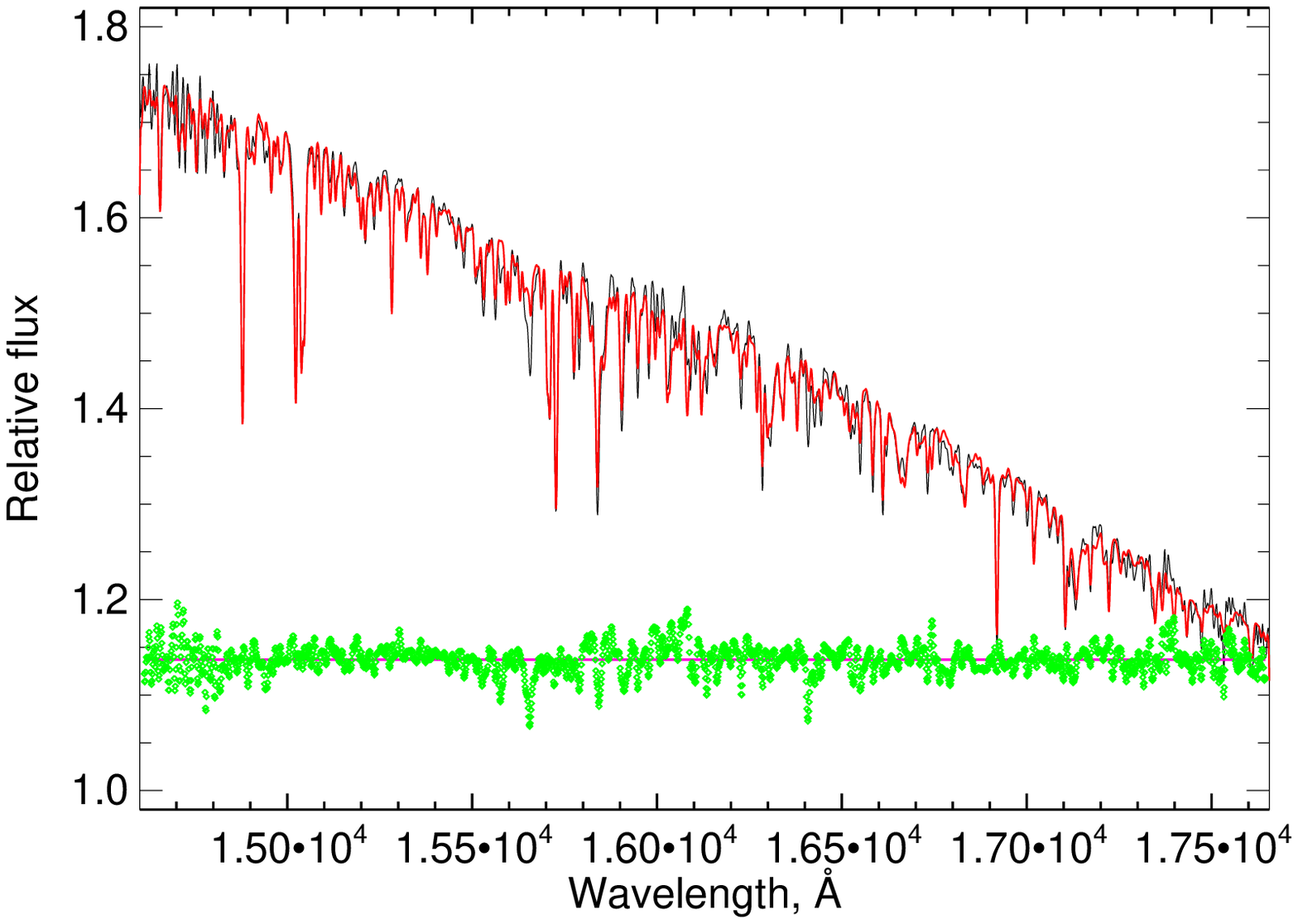}
	\includegraphics[width=0.925\columnwidth]{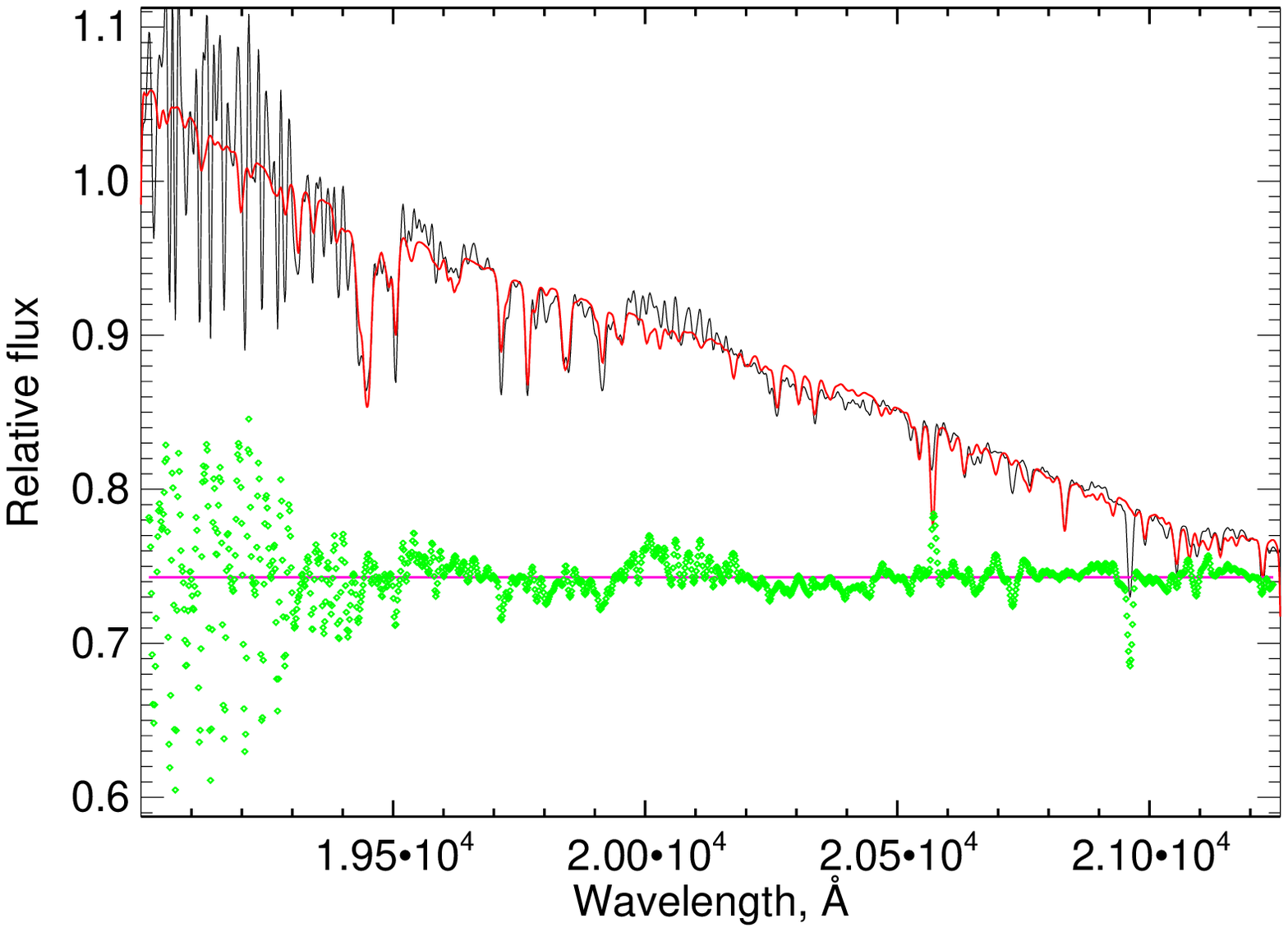}
	\includegraphics[width=0.925\columnwidth]{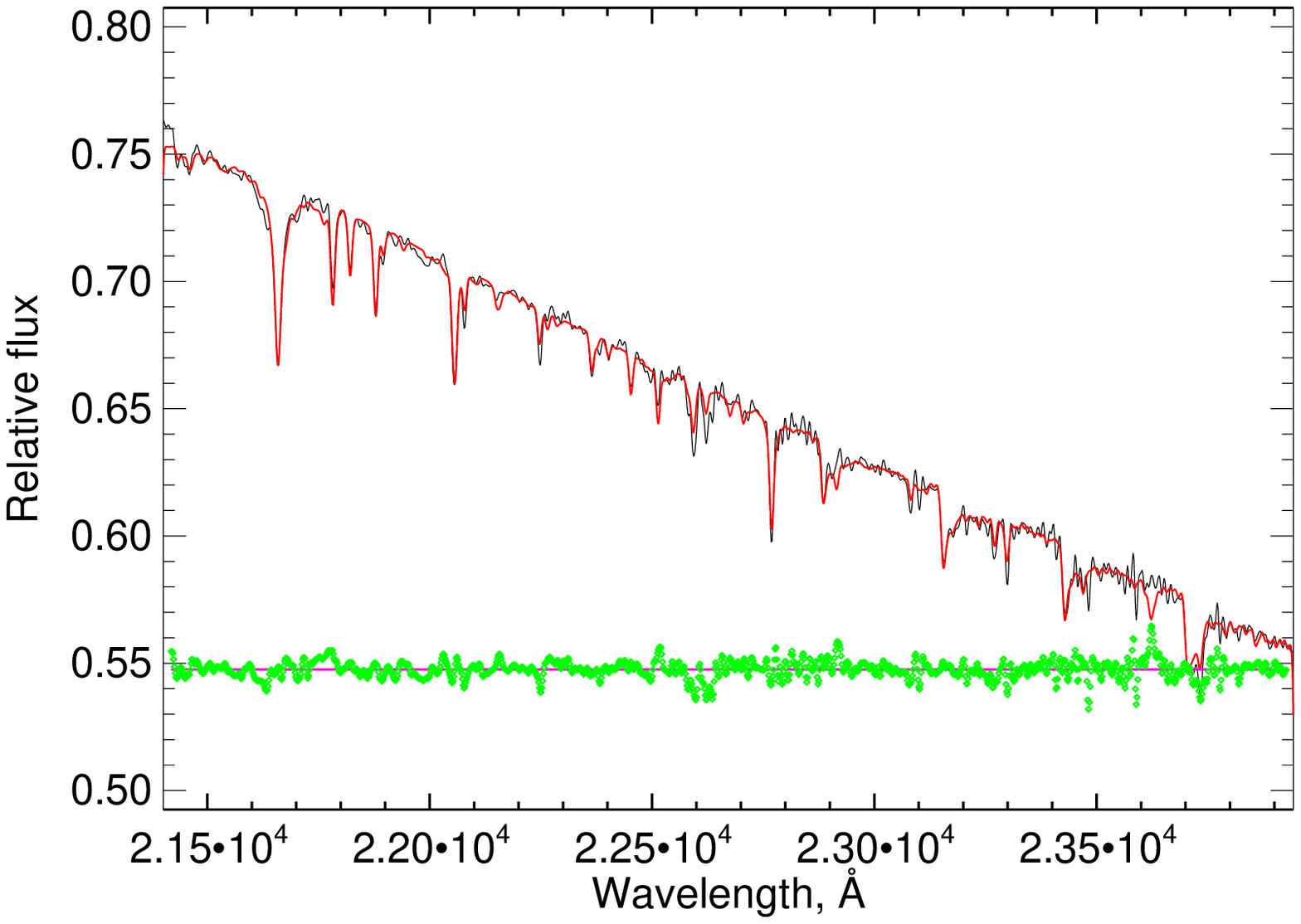}
	\caption{Example of the full-spectrum fitting for a solar-like star IRL075. We show the observed spectrum (black lines), the best fit model (red lines) and the residuals (green data points). In some regions of the fit is poor is due to telluric absorption contamination.}
	\label{irtf_fitting_fwhm}
\end{figure*}

\subsection{Comparison with recent work}
\label{irtf_final_compare}
\hspace{0.45cm}
We compare the atmospheric parameters of the  {\it IRTF spectral library} stars with the survey of available stellar parameters compiled from the literature by \citet{cesetti_et_al_2013}. In Figure \,\ref{irtf_cesetti} we present the residuals of the comparisons for $T_{\mathrm{eff}}$ ($184$ stars), $\log g$ ($101$ stars) and $[Z/Z_{\odot}]$ (95 stars). This comparison shows a reasonable agreement between our anchoring literature values from C07, our obtained ones and those from \citet{cesetti_et_al_2013}, with $\sigma_{T_{eff}} = 195 K$, $\sigma_{\log g}= 0.32\,\mathrm{dex}$ and $\sigma_{[Z/Z_{\odot}]}= 0.19\,\mathrm{dex}$.


\section{Flux calibration of the IRTF spectral library}
\label{irtf_tests}
\hspace{0.45cm}

In this section, we test the flux calibration of the library and the behaviour of the stellar integrated colours as a function of the atmospheric parameters. The reliability of this calibration is important for the SP modeling when comparing to photometric observations of globular clusters and galaxies.

Relative $J$, $H$ and $K_{s}$ fluxes were determined by integrating the spectral flux in these Near-IR  bands using the Vega spectrum from \citet{colina_et_al_1996} as a zero-point. We use the response curves of the $J$, $H$ and $K$ filters of the Johnson-Cousins-Glass photometric system given by \citet{bessel_et_al_1998}. We present in Figure \,\ref{irtf_teff_colors} a comparison of the colours with the library of \citet{pickles_1998} for solar metallicity stars. We can see that our inferred parameters follow the same trend as those of Pickles. We draw the same conclusion from the colour-colour diagrams in Figure \,\ref{irtf_color_color}.
 

\section{Determining the spectral resolution}
\label{fwhm}
\hspace{0.45cm}   

In order to asses the accuracy of our SSP models (Paper II), it is also important to characterise the spectral resolution of the stellar library to be used and to make sure that the radial velocities of all the stars have been set to zero. Therefore, to determine the nominal spectral resolution (FWHM) of the {\it IRTF spectral library}, we fit the stars with another template library of very high resolution.

We use the recent PHOENIX BT-Settl\footnote{\url{phoenix.ens-lyon.fr/Grids/BT-Settl/AGSS2009/SPECTRA/}} stellar models \citep{allard_et_al_2012} as a template library. The stellar parameters of this theoretical library cover a wide range of temperature (2600 $-7000\,\mathrm{K}$), luminosity class (from dwarfs to supergiants) and metallicity ($-0.5$ to $0.5\,\mathrm{dex}$). In the wavelength range that we use ($J$, $H$ and $K$ bands), the full-width at half maximum (FWHM) of the library is $0.05~\AA$. 

\begin{figure*}[!htb]
	\centering
	\includegraphics[width=\columnwidth]{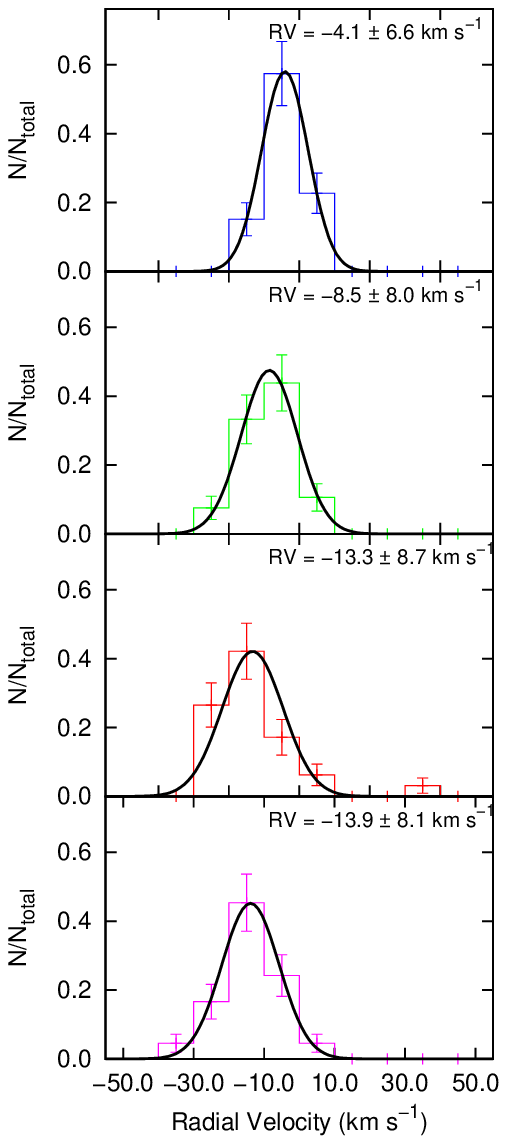}
	\includegraphics[width=\columnwidth]{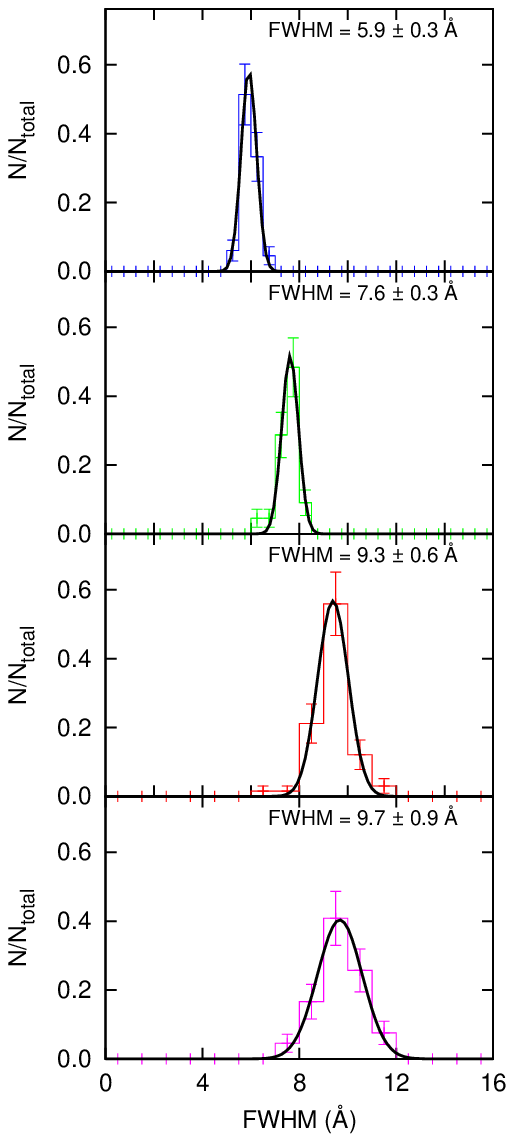}
	\caption{Left panels: Radial velocities of the {\it IRTF spectral library} as a function of wavelength. Right panel: FWHM as a function of wavelength. For the $J$ band, at $1.22~\mu m$, the stars have an average FWHM of $5.9~\pm 0.3~\AA$. In the $H$ band, at $1.62~\mu m$, the peak value is $7.6~\pm 0.3~\AA$. For the {\it atomic}-dominated part of the $K$ band, at $2.02~\mu m$, the average resolution is $9.3~\pm 0.6~\AA$. And for the {\it molecular}-dominated part of the $K$ band, at $2.27~\mu m$, the average FWHM is $9.7~\pm 0.9~\AA$.}
	\label{irtf_hists}
\end{figure*}

\begin{figure*}[!htb]
	\centering
	\includegraphics[width=\columnwidth]{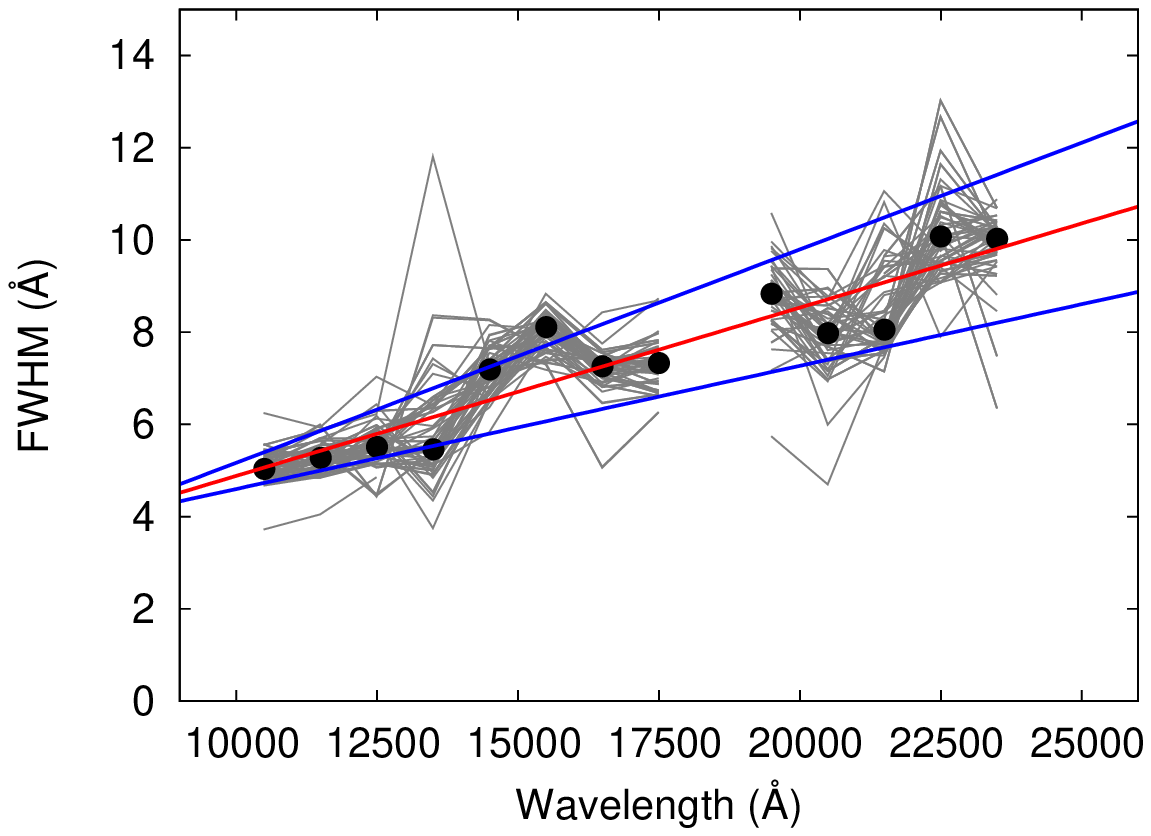}
	\includegraphics[width=\columnwidth]{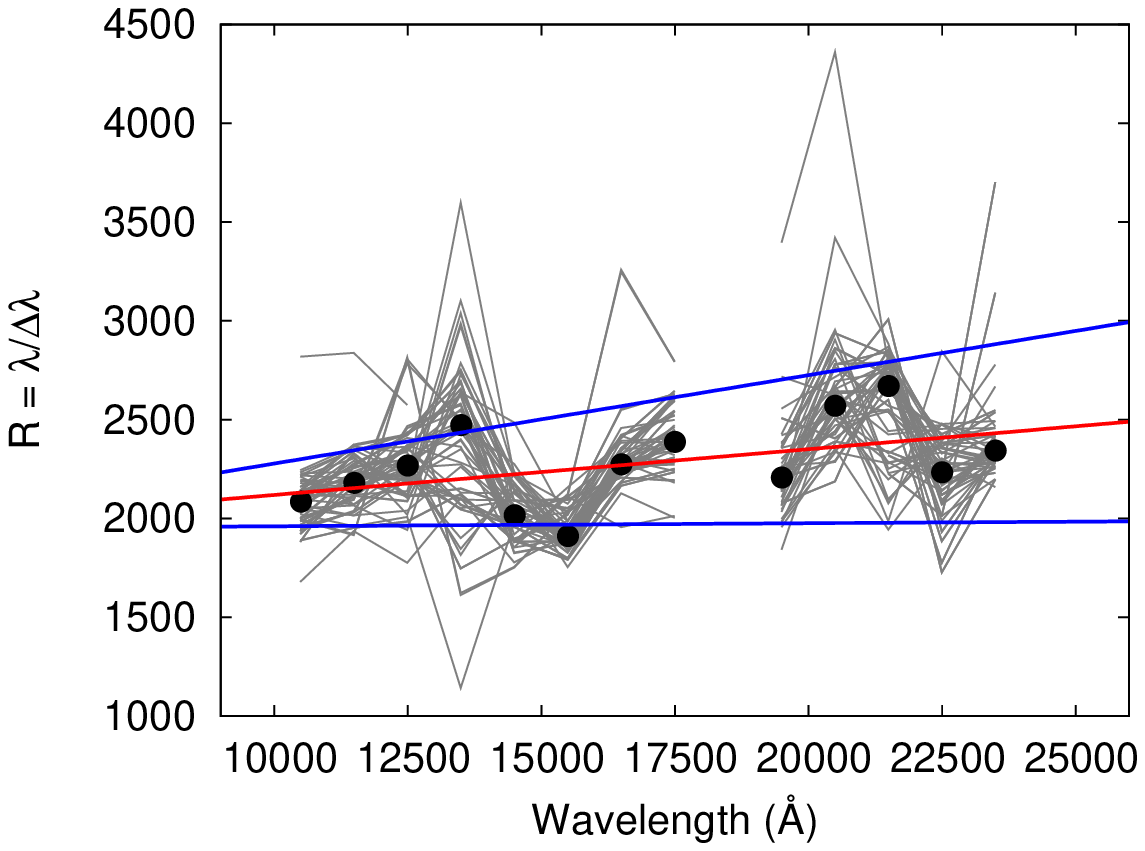}
	\caption{Behaviour of the FWHM (left panel) and the resolving power $R$ (right panel) of the G stars of the {\it IRTF spectral library} (grey lines) as a function of wavelength. In both panels, the black points represent the mean values for those effective wavelengths and the blue lines mark the mean dispersion. In the left panel, the red line corresponds to a liner relation of the mean FWHM for each wavelength. In the right panel, the red line is a linear relation of the mean $R$ with effective wavelength.}
	\label{irtf_fwhm_stats}
\end{figure*}

The FWHM of the {\it IRTF spectral library} has been determined by comparison with the BT$-$Settl library. Both libraries are rebinned to a velocity scale of 25~$\mathrm{km\,s^{-1}}$, after which the broadening of data and templates are measured using the penalized pixel-fitting method \citep[pPXF,][]{cappellari_and_emsellem_2004}. The templates have been convolved with a Gaussian and fit to each individual test spectrum to determine the FWHM. To assess the dependence of resolution on wavelength, we divide the original datasets into four wavelength bins, representing the $J$ ($1.04-1.44~\mu m$) and $H$ ($1.46-1.80~\mu m$) bands, the {\it atomic} part of the $K$ band ($1.90-2.14~\mu m$) and the {\it molecular} part of the $K$ band ($2.14-2.41~\mu m$), with effective wavelengths of $1.22$, $1.62$, $2.02$ and $2.27~\mu m$, respectively. The basic procedure is described in detail by \citet{falcon-barroso_et_al_2011}. Figure \,\ref{irtf_fitting_fwhm} shows an example of a solar-like star of the IRTF library and its fit using this procedure.

From this technique, we also obtain the radial velocity (RV) of the IRTF stars for each bin as shown in the left panel of Figure\,\ref{irtf_hists}. For the $J$ band, at $1.22~\mu m$, most of stars have an RV of $-4.1 \pm 6.6~\mathrm{km\,s^{-1}}$ (standard deviation). In the $H$ band, at $1.62~\AA$, the peak value is $-$8.5 $\pm 8.0~\mathrm{km\,s^{-1}}$. For the {\it atomic}-dominated part of the $K$ band, at $2.02~\AA$, the radial velocity is $-$13.3 $\pm 8.7~\mathrm{km\,s^{-1}}$, and for the {\it molecular}-dominated part of the $K$ band, at $2.27~\AA$, the RV is $-$13.9 $\pm 8.1~\mathrm{km\,s^{-1}}$.

In Figure \,\ref{irtf_fwhm_stats} we show how the FWHM, the resolution $R$ and their respective scatters behave as a function of wavelength for the G stars of the library, in bin sizes of $1000~\AA$. As we see, the FWHM increases when going to the red.

The behaviour of the FWHM as a function of each wavelength bin is presented in the right panel histograms of Figure \,\ref{irtf_hists}. One can see that the resolution is not constant in wavelength, as is the case for the MILES library, but is approximately proportional with wavelength ($R$ constant), but with some additional variation as a function of wavelength. For the $J$ band, at $1.22~\mu m$, most stars have an average FWHM of $5.9~\pm 0.3~\AA$. In the $H$ band, at $1.62~\AA$, the peak value is $7.6~\pm 0.3~\AA$. For the {\it atomic}-dominated part of the $K$ band, at $2.02~\AA$, the average resolution is $9.3~\pm 0.6~\AA$. And for the {\it molecular}-dominated part of the $K$ band, at $2.27~\AA$, the average FWHM is $9.7~\pm 0.9~\AA$. In terms of the spectral resolution, $R=2060 \pm 15$, $2163 \pm 16$, $2153 \pm 23$, and $2350 \pm 25$ at 1.22, 1.62, 2.02, and $2.27 \,\mu\mathrm{m}$ respectively.

These tests demonstrate that the resolution of this library is higher than most previous empirical libraries, such as \citet{lancon_and_wood_2000}.


\section{Final remarks}
\hspace{0.45cm}

We have studied in detail the accuracy and characteristics of the {\it IRTF spectral library} in the $J$, $H$ and $K$ bands. We have determined the the parameter coverage of the {\it IRTF spectral library}, allowing us to understand the extent of its usefulness. We have also determined the accuracy of the flux calibration of the library. Additionally, we have measured the precise spectral resolution and radial velocity of the stars. 

With these tests, we understand the possibilities and limitations of stellar population models using the {\it IRTF spectral library} as input (Paper II), over the $J$, $H$ and $K$ bands.

Cool late-type stars have a strong influence on the integrated flux in these wavelength regions. These stars are particularly relevant for studies of early-type galaxies. Since our library has cool stars, we can obtain a much better handle on the relative contribution of these stars in unresolved galaxies (Paper III). 

Our rebinned {\it IRTF spectral library} spectra (i.e. with a constant dispersion in wavelength) and the resulting data of the analysis and characterisation are available on-line\footnote{\url{smg.astro-research.net/}}


\section*{Acknowledgments}
\hspace{0.45cm}

The authors acknowledge the usage of the SIMBAD data base (operated at CDS, Strasbourg, France) and the {\it IRTF spectral library} database. SMG thanks T. de Boer and the Institute of Astronomy of the University of Cambridge for support during her stay. MK is a postdoctoral Marie Curie fellow (Grant PIEF-GA-2010-271780) and a fellow of the Fund for Scientific Research - Flanders, Belgium ($FWO11/PDO/147$). JFB and AV acknowledge the support by the Programa Nacional del Astronom{\'{i}}a y Astrof{\'{i}}sica of the Spanish Ministry of Science and Innovation undergrant $AYA2013-48226-C3-1-P$, as well as from the FP7 Marie Curie Actions of the European Commission, via the Initial Training Network DAGAL under REA grant agreement number $289313$.


\bibliographystyle{aa}
\bibliography{references}


\appendix
\onecolumn

\section[]{The stars composing the {\it IRTF spectral library} and their atmospheric parameters.}

In this appendix, we compile the available information for the IRTF stars. The first column shows the ID of the IRTF stars, corresponding to the order of stars adopted in \citet{rayner_et_al_2009, cushing_et_al_2005}. Indented IDs indicate stars which are not corrected for extinction. The second column lists the names given by the IRTF library database. The coordinates of each star are given in the third and fourth column. The fifth column gives the stellar class of each star according also to the IRTF library. The fifth, sixth and seventh columns present the atmospheric parameters of each star, determined from either the full-spectrum fitting method ($^{FSF}$), the colour$-$temperature relation ($^{CTR}$), literature compilations S10 and C97 ($^{SEC}$) or according to their stellar type ($^{ST}$). For stars from the \citet[][]{cenarro_et_al_2007} catalog~(labeled $^{C07}$) the atmospheric parameters are determined using the full-spectrum fitting method and recovered as the same values as given in the literature. Finally, the eight, ninth and tenth columns list the NIR colours of each star as calculated from the spectrum~(see Section~\ref{irtf_tests}). 

\begin{footnotesize}
\begin{longtable}{lllllllllll}
	\caption{Reference relation for the {\it IRTF spectral library} stars.}
	\label{nomenclature}	\\
	\hline
	\hline
	{\bf ID} & {\bf Star} & {\bf RA} & {\bf Dec} & {\bf Class} & {\bf $T_{\mathrm{eff}}$ (K)} & {\bf $\log g$} & {\bf $[Z/Z_{\odot}]$ } & (J-H) & (J-K) & (H-K) \\
	\hline
	\hline
	\endfirsthead
	\hline
	\hline
	\endlastfoot
	\multicolumn{11}{c}{\tablename\ \thetable\ - Continued from previous page} \\
	\multicolumn{11}{c}{ } \\
	\hline
	\hline
	{\bf ID} & {\bf Star} & {\bf RA} & {\bf Dec} & {\bf Class} & {\bf $T_{\mathrm{eff}}$ (K)} & {\bf $\log g$} & {\bf $[Z/Z_{\odot}]$ } & (J-H) & (J-K) & (H-K) \\
	\hline
	\hline
	\endhead
	\hline
	\hline
	\endfoot
IRL280     &  BRI B0021$-$0214		  &  00 24 24.63   &  $-$01 58 20.14  &  M9.5V      &  2466$^{CTR}$ & 5.06$^{ST }$  &  0.00$^{ST }$  & 0.81 & 1.36 & 0.55  \\
IRL147     &  HD 002901 		  &  00 32 47.52   &  +54 07 11.81    &  K2III      &  4410$^{FSF}$ & 2.68$^{ST }$  & -0.07$^{FSF}$  & 0.67 & 0.75 & 0.08  \\
IRL194     &  2MASS J00361617+1821104	  &  00 36 16.17   &  +18 21 10.47    &  L3.5       &  2324$^{CTR}$ & $\ge$5.0$^{ST }$  &  0.00$^{ST }$  & 0.90 & 1.45 & 0.56  \\
IRL181     &  HD 003346 		  &  00 36 46.44   &  +44 29 18.91    &  K6IIIa     &  3943$^{FSF}$ & 2.13$^{ST }$  & -0.05$^{FSF}$  & 0.84 & 1.05 & 0.21  \\
IRL072     &  HD 003421 		  &  00 37 21.21   &  +35 23 58.20    &  G2Ib       &  5663$^{FSF}$ & 1.94$^{FSF}$  & -0.33$^{FSF}$  & 0.42 & 0.48 & 0.06  \\
IRL153     &  HD 003765 		  &  00 40 49.26   &  +40 11 13.83    &  K2V        &  5073$^{FSF}$ & 4.49$^{FSF}$  &  0.10$^{FSF}$  & 0.46 & 0.53 & 0.07  \\
IRL252     &  HD 004408 		  &  00 46 32.95   &  +15 28 31.81    &  M4III      &  3566$^{CTR}$ & 1.07$^{ST }$  &  0.00$^{ST }$  & 0.93 & 1.21 & 0.27  \\
IRL260     &  Gl 051			  &  01 03 19.72   &  +62 21 55.70    &  M5V        &  3649$^{CTR}$ & 4.8 $^{ST }$  &  0.00$^{ST }$  & 0.61 & 0.92 & 0.31  \\
IRL014     &  HD 006130 		  &  01 03 37.00   &  +61 04 29.36    &  F0II       &  7031$^{FSF}$ & 2.15$^{FSF}$  & -0.55$^{FSF}$  & 0.25 & 0.33 & 0.08  \\
\hspace{0.2cm}IRL013     &  HD 006130 ext		  &  01 03 37.00   &  +61 04 29.36    &  F0II       &  7008$^{FSF}$ & 2.09$^{FSF}$  & -0.57$^{FSF}$  & 0.16 & 0.19 & 0.03  \\
IRL277     &  IRAS 01037+1219		  &  01 06 25.98   &  +12 35 53.00    &  M8III      &  3777$^{CTR}$ & 0.7 $^{ST }$  &  0.00$^{ST }$  & 2.50 & 4.21 & 1.71  \\
\hspace{0.2cm}IRL082     &  HD 006474 ext		  &  01 06 59.74   &  +63 46 23.38    &  G4Ia       &  6241$^{C07}$ & 1.55$^{C07}$  &  0.25$^{C07}$  & 0.35 & 0.46 & 0.10  \\
IRL083     &  HD 006474 		  &  01 06 59.74   &  +63 46 23.38    &  G4Ia       &  6241$^{C07}$ & 1.55$^{C07}$  &  0.25$^{C07}$  & 0.55 & 0.77 & 0.22  \\
IRL055     &  HD 006903 		  &  01 09 49.20   &  +19 39 30.26    &  F9IIIa     &  5570$^{C07}$ & 2.9 $^{C07}$  & -0.35$^{C07}$  & 0.36 & 0.42 & 0.06  \\
\hspace{0.2cm}IRL054     &  HD 006903 ext		  &  01 09 49.20   &  +19 39 30.26    &  F9IIIa     &  5570$^{C07}$ & 2.9 $^{C07}$  & -0.35$^{C07}$  & 0.32 & 0.37 & 0.04  \\
IRL202     &  HD 236697 		  &  01 19 53.61   &  +58 18 30.73    &  M0.5Ib     &  3565$^{CTR}$ & 0.69$^{ST }$  &  0.00$^{ST }$  & 0.93 & 1.21 & 0.28  \\
\hspace{0.2cm}IRL201     &  HD 236697 ext		  &  01 19 53.61   &  +58 18 30.73    &  M0.5Ib     &  3814$^{CTR}$ & 0.69$^{ST }$  &  0.00$^{ST }$  & 0.78 & 0.97 & 0.19  \\
IRL010     &  HD 007927 		  &  01 20 04.91   &  +58 13 53.80    &  F0Ia       &  7425$^{C07}$ & 0.7 $^{C07}$  &  0.00$^{C07}$  & 0.29 & 0.43 & 0.14  \\
\hspace{0.2cm}IRL009     &  HD 007927 ext		  &  01 20 04.91   &  +58 13 53.80    &  F0Ia       &  7425$^{C07}$ & 0.7 $^{C07}$  &  0.00$^{C07}$  & 0.13 & 0.17 & 0.04  \\
\hspace{0.2cm}IRL209     &  BD+60265 ext		  &  01 33 33.05   &  +61 33 30.73    &  M1.5Ib     &  3771$^{CTR}$ & 0.61$^{ST }$  &  0.00$^{ST }$  & 0.76 & 1.01 & 0.24  \\
IRL210     &  BD+60265  		  &  01 33 33.05   &  +61 33 30.73    &  M1.5Ib     &  3468$^{CTR}$ & 0.61$^{ST }$  &  0.00$^{ST }$  & 0.97 & 1.33 & 0.36  \\
IRL124     &  HD 009852 		  &  01 37 51.23   &  +61 51 41.74    &  K0.5III    &  4678$^{FSF}$ & 2.88$^{ST }$  &  0.01$^{FSF}$  & 0.67 & 0.81 & 0.14  \\
\hspace{0.2cm}IRL123     &  HD 009852 ext		  &  01 37 51.23   &  +61 51 41.74    &  K0.5III    &  4740$^{FSF}$ & 2.88$^{ST }$  & -0.05$^{FSF}$  & 0.55 & 0.61 & 0.06  \\
IRL067     &  HD 010307 		  &  01 41 47.14   &  +42 36 48.12    &  G1V        &  5847$^{C07}$ & 4.28$^{C07}$  &  0.02$^{C07}$  & 0.32 & 0.34 & 0.03  \\
IRL144     &  HD 010476 		  &  01 42 29.76   &  +20 16 06.60    &  K1V        &  5150$^{C07}$ & 4.44$^{C07}$  & -0.17$^{C07}$  & 0.45 & 0.53 & 0.08  \\
\hspace{0.2cm}IRL225     &  HD 010465 ext		  &  01 43 11.10   &  +48 31 00.36    &  M2Ib       &  3783$^{CTR}$ & 0.62$^{ST }$  &  0.00$^{ST }$  & 0.78 & 1.00 & 0.22  \\
IRL226     &  HD 010465 		  &  01 43 11.10   &  +48 31 00.36    &  M2Ib       &  3673$^{CTR}$ & 0.62$^{ST }$  &  0.00$^{ST }$  & 0.84 & 1.09 & 0.26  \\
IRL081     &  HD 010697 		  &  01 44 55.82   &  +20 04 59.33    &  G3Va       &  5688$^{FSF}$ & 3.82$^{FSF}$  & -0.09$^{FSF}$  & 0.36 & 0.44 & 0.08  \\
IRL039     &  HD 011443 		  &  01 53 04.90   &  +29 34 43.78    &  F6IV       &  6288$^{C07}$ & 3.91$^{C07}$  &  0.00$^{C07}$  & 0.27 & 0.35 & 0.08  \\
IRL189     &  2MASS J02081833+2542533	  &  02 08 18.33   &  +25 42 53.30    &  L1         &  2285$^{CTR}$ & $\ge$5.0$^{ST }$  &  0.00$^{ST }$  & 0.90 & 1.49 & 0.58  \\
IRL016     &  HD 013174 		  &  02 09 25.33   &  +25 56 23.59    &  F0III      &  7000$^{C07}$ & 3.25$^{C07}$  & -1.90$^{C07}$  & 0.18 & 0.19 & 0.01  \\
IRL257     &  HD 014386 		  &  02 19 20.79   &  $-$02 58 39.49  &  M5III      &  3668$^{CTR}$ & 1.04$^{ST }$  &  0.00$^{ST }$  & 0.63 & 1.10 & 0.47  \\
\hspace{0.2cm}IRL212     &  HD 014404 ext		  &  02 21 42.41   &  +57 51 46.12    &  M1Iab      &  3952$^{CTR}$ & 0.0 $^{SEC}$  &  0.00$^{ST }$  & 0.71 & 0.87 & 0.17  \\
IRL213     &  HD 014404 		  &  02 21 42.41   &  +57 51 46.12    &  M1Iab      &  3593$^{CTR}$ & 0.0 $^{SEC}$  &  0.00$^{ST }$  & 0.90 & 1.18 & 0.28  \\
\hspace{0.2cm}IRL242     &  HD 014469 ext		  &  02 22 06.89   &  +56 36 14.86    &  M3Iab      &  3773$^{CTR}$ & 0.16$^{ST }$  &  0.00$^{ST }$  & 0.75 & 1.00 & 0.25  \\
IRL243     &  HD 014469 		  &  02 22 06.89   &  +56 36 14.86    &  M3Iab      &  3538$^{CTR}$ & 0.16$^{ST }$  &  0.00$^{ST }$  & 0.90 & 1.24 & 0.34  \\
\hspace{0.2cm}IRL232     &  HD 014488 ext		  &  02 22 24.29   &  +57 06 34.36    &  M3.5Iab    &  3525$^{CTR}$ & 0.14$^{ST }$  &  0.00$^{ST }$  & 0.88 & 1.26 & 0.37  \\
IRL233     &  HD 014488 		  &  02 22 24.29   &  +57 06 34.36    &  M3.5Iab    &  3340$^{CTR}$ & 0.14$^{ST }$  &  0.00$^{ST }$  & 1.05 & 1.51 & 0.47  \\
\hspace{0.2cm}IRL101     &  HD 016139 ext		  &  02 36 17.99   &  +27 28 20.36    &  G7.5IIIa   &  5123$^{FSF}$ & 2.88$^{FSF}$  & -0.31$^{FSF}$  & 0.47 & 0.57 & 0.10  \\
IRL102     &  HD 016139 		  &  02 36 17.99   &  +27 28 20.36    &  G7.5IIIa   &  5069$^{FSF}$ & 2.7 $^{FSF}$  & -0.41$^{FSF}$  & 0.51 & 0.63 & 0.12  \\
IRL166     &  HD 016068 		  &  02 36 52.80   &  +55 54 55.41    &  K3II       &  4285$^{FSF}$ & 1.88$^{FSF}$  & -0.09$^{FSF}$  & 0.78 & 0.94 & 0.15  \\
\hspace{0.2cm}IRL165     &  HD 016068 ext		  &  02 36 52.80   &  +55 54 55.41    &  K3II       &  4380$^{FSF}$ & 1.73$^{FSF}$  & -0.06$^{FSF}$  & 0.62 & 0.69 & 0.06  \\
IRL029     &  HD 016232 		  &  02 36 57.74   &  +24 38 53.02    &  F4V        &  6097$^{FSF}$ & 4.08$^{FSF}$  & -0.02$^{FSF}$  & 0.23 & 0.26 & 0.03  \\
IRL034     &  HD 017918 		  &  02 53 11.69   &  +16 29 00.45    &  F5III      &  6695$^{FSF}$ & 3.25$^{FSF}$  &  0.00$^{ST }$  & 0.22 & 0.22 & 0.01  \\
IRL200     &  DENIS$-$P J0255.0$-$4700	  &  02 55 03.57   &  $-$47 00 50.99  &  L8         &  1952$^{CTR}$ & $\ge$5.0$^{ST }$  &  0.00$^{ST }$  & 1.10 & 1.72 & 0.61  \\
IRL266     &  HD 018191 		  &  02 55 48.49   &  +18 19 53.90    &  M6III      &  3536$^{CTR}$ & 1.92$^{C07}$  & -0.99$^{C07}$  & 0.95 & 1.24 & 0.29  \\
IRL092     &  HD 018474 		  &  02 59 49.79   &  +47 13 14.48    &  G5III      &  5878$^{FSF}$ & 3.73$^{FSF}$  & -0.11$^{FSF}$  & 0.46 & 0.50 & 0.03  \\
IRL249     &  HD 019058 		  &  03 05 10.59   &  +38 50 24.99    &  M4IIIa     &  3549$^{CTR}$ & 0.57$^{SEC}$  & -0.22$^{SEC}$  & 0.96 & 1.23 & 0.27  \\
IRL205     &  HD 019305 		  &  03 06 26.73   &  +01 57 54.63    &  M0V        &  3934$^{CTR}$ & 4.65$^{ST }$  &  0.00$^{ST }$  & 0.69 & 0.85 & 0.16  \\
IRL061     &  HD 020619 		  &  03 19 01.89   &  $-$02 50 35.49  &  G1.5V      &  6094$^{FSF}$ & 4.04$^{FSF}$  & -0.18$^{FSF}$  & 0.38 & 0.45 & 0.07  \\
IRL109     &  HD 020618 		  &  03 19 55.79   &  +27 04 16.06    &  G7IV       &  5430$^{FSF}$ & 3.16$^{FSF}$  & -0.25$^{FSF}$  & 0.49 & 0.57 & 0.09  \\
IRL279     &  LP 412$-$31 		  &  03 20 59.65   &  +18 54 23.31    &  M8V        &  2841$^{CTR}$ & 4.96$^{ST }$  &  0.00$^{ST }$  & 0.72 & 1.17 & 0.45  \\
\hspace{0.2cm}IRL063     &  HD 021018 ext		  &  03 23 38.98   &  +04 52 55.57    &  G1III      &  5924$^{FSF}$ & 3.48$^{FSF}$  & -0.04$^{FSF}$  & 0.35 & 0.41 & 0.06  \\
IRL064     &  HD 021018 		  &  03 23 38.98   &  +04 52 55.57    &  G1III      &  5920$^{FSF}$ & 3.58$^{FSF}$  & -0.04$^{FSF}$  & 0.40 & 0.49 & 0.09  \\
IRL028     &  HD 021770 		  &  03 32 26.26   &  +46 03 24.69    &  F4III      &  6204$^{FSF}$ & 3.75$^{ST }$  & -0.04$^{SEC}$  & 0.21 & 0.23 & 0.02  \\
IRL286     &  LP 944$-$20 		  &  03 39 35.22   &  $-$35 25 44.09  &  M9V        &  2821$^{CTR}$ & 5.0 $^{ST }$  &  0.00$^{ST }$  & 0.70 & 1.20 & 0.50  \\
\hspace{0.2cm}IRL145     &  HD 023082 ext		  &  03 44 05.77   &  +44 53 04.91    &  K2.5II     &  4270$^{FSF}$ & 1.22$^{FSF}$  &  0.00$^{FSF}$  & 0.63 & 0.73 & 0.10  \\
IRL146     &  HD 023082 		  &  03 44 05.77   &  +44 53 04.91    &  K2.5II     &  4215$^{FSF}$ & 1.42$^{FSF}$  & -0.16$^{FSF}$  & 0.78 & 0.96 & 0.18  \\
\hspace{0.2cm}IRL227     &  HD 023475 ext		  &  03 49 31.27   &  +65 31 33.50    &  M2II       &  3800$^{CTR}$ & 1.0 $^{ST }$  &  0.00$^{ST }$  & 0.80 & 0.98 & 0.18  \\
IRL228     &  HD 023475 		  &  03 49 31.27   &  +65 31 33.50    &  M2II       &  3650$^{CTR}$ & 1.0 $^{ST }$  &  0.00$^{ST }$  & 0.88 & 1.12 & 0.23  \\
IRL027     &  HD 026015 		  &  04 07 41.98   &  +15 09 46.02    &  F3V        &  7090$^{CTR}$ & 4.29$^{ST }$  &  0.04$^{SEC}$  & 0.14 & 0.12 & -0.02 \\
IRL141     &  HD 025975 		  &  04 08 15.38   &  +37 43 38.98    &  K1III      &  4966$^{FSF}$ & 2.92$^{FSF}$  & -0.24$^{FSF}$  & 0.47 & 0.53 & 0.06  \\
IRL106     &  HD 025877 		  &  04 09 27.57   &  +59 54 29.05    &  G7II       &  4904$^{C07}$ & 1.3 $^{C07}$  & -0.15$^{C07}$  & 0.45 & 0.52 & 0.06  \\
\hspace{0.2cm}IRL105     &  HD 025877 ext		  &  04 09 27.57   &  +59 54 29.05    &  G7II       &  4904$^{C07}$ & 1.3 $^{C07}$  & -0.15$^{C07}$  & 0.39 & 0.42 & 0.03  \\
IRL052     &  HD 027383 		  &  04 19 54.85   &  +16 31 21.32    &  F8V        &  6127$^{FSF}$ & 4.02$^{FSF}$  & -0.05$^{FSF}$  & 0.24 & 0.25 & 0.02  \\
IRL017     &  HD 027397 		  &  04 19 57.70   &  +14 02 06.71    &  F0IV       &  6799$^{FSF}$ & 4.1 $^{ST }$  &  0.00$^{ST }$  & 0.13 & 0.11 & -0.02 \\
IRL251     &  HD 027598 		  &  04 20 41.34   &  $-$16 49 47.91  &  M4III      &  3623$^{CTR}$ & 1.07$^{ST }$  &  0.00$^{ST }$  & 0.90 & 1.15 & 0.25  \\
IRL100     &  HD 027277 		  &  04 20 53.54   &  +50 16 18.20    &  G6III      &  5196$^{FSF}$ & 2.54$^{FSF}$  & -0.20$^{FSF}$  & 0.47 & 0.55 & 0.08  \\
IRL037     &  HD 027524 		  &  04 21 31.64   &  +21 02 23.56    &  F5V        &  6576$^{C07}$ & 4.25$^{C07}$  &  0.13$^{C07}$  & 0.21 & 0.23 & 0.02  \\
IRL235     &  HD 028487 		  &  04 29 38.94   &  +05 09 51.35    &  M3.5III    &  3553$^{CTR}$ & 1.27$^{ST }$  &  0.00$^{ST }$  & 0.96 & 1.22 & 0.27  \\
\hspace{0.2cm}IRL234     &  HD 028487 ext		  &  04 29 38.94   &  +05 09 51.35    &  M3.5III    &  3603$^{CTR}$ & 1.27$^{ST }$  &  0.00$^{ST }$  & 0.92 & 1.17 & 0.25  \\
IRL001     &  HD 031996 		  &  04 59 36.34   &  $-$14 48 22.53  &  C7.6       &  2181$^{CTR}$ & -0.5$^{C07}$  &  0.20$^{C07}$  & 1.40 & 2.55 & 1.14  \\
\hspace{0.2cm}IRL207     &  HD 035601 ext		  &  05 27 10.21   &  +29 55 15.79    &  M1.5Iab    &  3738$^{CTR}$ & 0.0 $^{C07}$  & -0.20$^{C07}$  & 0.79 & 1.03 & 0.24  \\
IRL208     &  HD 035601 		  &  05 27 10.21   &  +29 55 15.79    &  M1.5Iab    &  3497$^{CTR}$ & 0.0 $^{C07}$  & -0.20$^{C07}$  & 0.95 & 1.29 & 0.34  \\
\hspace{0.2cm}IRL159     &  HD 035620 ext		  &  05 27 38.88   &  +34 28 33.21    &  K3III      &  4367$^{C07}$ & 1.75$^{C07}$  & -0.03$^{C07}$  & 0.64 & 0.77 & 0.13  \\
IRL160     &  HD 035620 		  &  05 27 38.88   &  +34 28 33.21    &  K3III      &  4367$^{C07}$ & 1.75$^{C07}$  & -0.03$^{C07}$  & 0.68 & 0.84 & 0.15  \\
IRL180     &  HD 036003 		  &  05 28 26.09   &  $-$03 29 58.39  &  K5V        &  4464$^{C07}$ & 4.61$^{C07}$  &  0.09$^{C07}$  & 0.61 & 0.72 & 0.11  \\
IRL140     &  HD 036134 		  &  05 29 23.68   &  $-$03 26 47.01  &  K1III      &  4837$^{FSF}$ & 2.24$^{FSF}$  & -0.32$^{FSF}$  & 0.60 & 0.68 & 0.08  \\
IRL211     &  HD 036395 		  &  05 31 27.39   &  $-$03 40 38.03  &  M1.5V      &  3651$^{CTR}$ & 4.57$^{SEC}$  &  0.53$^{SEC}$  & 0.73 & 0.92 & 0.18  \\
IRL197     &  SDSS J053951.99$-$005902.0  &  05 39 52.00   &  $-$00 59 01.90  &  L5         &  2325$^{CTR}$ & $\ge$5.0$^{ST }$  &  0.00$^{ST }$  & 0.93 & 1.45 & 0.52  \\
IRL253     &  Gl 213			  &  05 42 09.26   &  +12 29 21.62    &  M4V        &  4063$^{CTR}$ & 4.75$^{ST }$  &  0.00$^{ST }$  & 0.53 & 0.79 & 0.26  \\
\hspace{0.2cm}IRL240     &  HD 039045 ext		  &  05 51 25.74   &  +32 07 28.89    &  M3III      &  3690$^{CTR}$ & 1.25$^{ST }$  &  0.00$^{ST }$  & 0.86 & 1.08 & 0.22  \\
IRL241     &  HD 039045 		  &  05 51 25.74   &  +32 07 28.89    &  M3III      &  3619$^{CTR}$ & 1.25$^{ST }$  &  0.00$^{ST }$  & 0.90 & 1.15 & 0.24  \\
\hspace{0.2cm}IRL217     &  HD 039801 ext		  &  05 55 10.30   &  +07 24 25.43    &  M2Ia       &  4190$^{CTR}$ & 0.0 $^{C07}$  &  0.05$^{C07}$  & 0.68 & 0.73 & 0.05  \\
IRL218     &  HD 039801 		  &  05 55 10.30   &  +07 24 25.43    &  M2Ia       &  4016$^{CTR}$ & 0.0 $^{C07}$  &  0.05$^{C07}$  & 0.74 & 0.83 & 0.09  \\
\hspace{0.2cm}IRL068     &  HD 039949 ext		  &  05 57 05.55   &  +27 18 59.95    &  G2Ib       &  5671$^{FSF}$ & 1.55$^{ST }$  & -0.07$^{FSF}$  & 0.41 & 0.48 & 0.07  \\
IRL069     &  HD 039949 		  &  05 57 05.55   &  +27 18 59.95    &  G2Ib       &  5726$^{FSF}$ & 1.55$^{ST }$  & -0.25$^{FSF}$  & 0.47 & 0.58 & 0.11  \\
IRL025     &  HD 040535 		  &  05 59 01.08   &  $-$09 22 56.00  &  F2III      &  6791$^{FSF}$ & 3.81$^{ST }$  & -0.24$^{FSF}$  & 0.21 & 0.25 & 0.04  \\
IRL292     &  2MASS J05591915$-$1404489	  &  05 59 19.14   &  $-$14 04 48.88  &  T4.5       &  1105$^{CTR}$ & $\ge$5.0$^{ST }$  &  0.67$^{SEC}$  & 0.21 & 0.09 & -0.12 \\
IRL239     &  HD 040239 		  &  05 59 56.09   &  +45 56 12.24    &  M3IIb      &  3558$^{CTR}$ & 1.0 $^{ST }$  &  0.00$^{ST }$  & 0.94 & 1.22 & 0.28  \\
IRL219     &  HD 042581 		  &  06 10 34.61   &  $-$21 51 52.71  &  M1V        &  3937$^{CTR}$ & 4.69$^{ST }$  &  0.00$^{ST }$  & 0.66 & 0.85 & 0.19  \\
\hspace{0.2cm}IRL070     &  HD 042454 ext		  &  06 12 05.48   &  +29 29 31.72    &  G2Ib       &  5772$^{FSF}$ & 1.55$^{ST }$  & -0.01$^{FSF}$  & 0.36 & 0.40 & 0.05  \\
IRL071     &  HD 042454 		  &  06 12 05.48   &  +29 29 31.72    &  G2Ib       &  5751$^{FSF}$ & 1.55$^{ST }$  &  0.02$^{FSF}$  & 0.46 & 0.57 & 0.11  \\
IRL290     &  HD 044544 		  &  06 22 23.85   &  +03 25 27.88    &  SC5.5      &  3055$^{CTR}$ & 0.2 $^{ST }$  & -1.12$^{SEC}$  & 1.15 & 1.53 & 0.38  \\
\hspace{0.2cm}IRL127     &  HD 044391 ext		  &  06 22 47.87   &  +27 59 12.03    &  K0Ib       &  4786$^{FSF}$ & 1.48$^{FSF}$  & -0.13$^{FSF}$  & 0.54 & 0.66 & 0.11  \\
IRL128     &  HD 044391 		  &  06 22 47.87   &  +27 59 12.03    &  K0Ib       &  4764$^{FSF}$ & 1.65$^{FSF}$  & -0.11$^{FSF}$  & 0.61 & 0.76 & 0.15  \\
IRL006     &  HD 044984 		  &  06 25 28.17   &  +14 43 19.16    &  CN4        &  3248$^{CTR}$ & -0.5$^{ST }$  & -0.17$^{SEC}$  & 1.01 & 1.41 & 0.40  \\
\hspace{0.2cm}IRL174     &  HD 045977 ext		  &  06 30 07.30   &  $-$11 48 32.19  &  K4V        &  4738$^{FSF}$ & 4.39$^{FSF}$  &  0.14$^{FSF}$  & 0.49 & 0.56 & 0.07  \\
IRL175     &  HD 045977 		  &  06 30 07.30   &  $-$11 48 32.19  &  K4V        &  4805$^{FSF}$ & 4.17$^{FSF}$  &  0.15$^{FSF}$  & 0.53 & 0.62 & 0.09  \\
IRL007     &  HD 048664 		  &  06 44 40.71   &  +03 18 58.65    &  CN5        &  2710$^{CTR}$ & -0.5$^{ST }$  & -0.62$^{SEC}$  & 1.17 & 1.78 & 0.61  \\
IRL048     &  HD 051956 		  &  06 59 31.74   &  +00 55 00.36    &  F8Ib       &  6130$^{FSF}$ & 1.7 $^{ST }$  & -0.17$^{FSF}$  & 0.35 & 0.42 & 0.07  \\
\hspace{0.2cm}IRL047     &  HD 051956 ext		  &  06 59 31.74   &  +00 55 00.36    &  F8Ib       &  6268$^{FSF}$ & 1.7 $^{ST }$  & -0.10$^{FSF}$  & 0.27 & 0.30 & 0.03  \\
IRL248     &  Gl 268			  &  07 10 01.83   &  +38 31 46.06    &  M4.5V      &  3772$^{CTR}$ & 4.75$^{ST }$  &  0.00$^{ST }$  & 0.60 & 0.89 & 0.29  \\
IRL003     &  HD 057160 		  &  07 20 59.00   &  +24 59 58.07    &  CJ5        &  2889$^{CTR}$ & -0.5$^{ST }$  &  0.00$^{ST }$  & 1.10 & 1.64 & 0.55  \\
IRL099     &  HD 058367 		  &  07 25 38.89   &  +09 16 33.95    &  G6IIb      &  5009$^{FSF}$ & 1.56$^{FSF}$  & -0.19$^{FSF}$  & 0.49 & 0.57 & 0.09  \\
IRL236     &  Gl 273			  &  07 27 24.49   &  +05 13 32.83    &  M3.5V      &  3971$^{CTR}$ & 4.74$^{ST }$  &  0.00$^{ST }$  & 0.58 & 0.84 & 0.26  \\
\hspace{0.2cm}IRL237     &  CD$-$314916 ext		  &  07 41 02.62   &  $-$31 40 59.10  &  M3Iab      &  3778$^{CTR}$ & 0.12$^{ST }$  &  0.00$^{ST }$  & 0.78 & 1.00 & 0.22  \\
IRL238     &  CD$-$314916 		  &  07 41 02.62   &  $-$31 40 59.10  &  M3Iab      &  3517$^{CTR}$ & 0.12$^{ST }$  &  0.00$^{ST }$  & 0.95 & 1.27 & 0.31  \\
IRL289     &  HD 062164 		  &  07 42 17.46   &  $-$10 52 47.18  &  S5         &  3198$^{CTR}$ & 0.45$^{ST }$  &  0.00$^{ST }$  & 1.06 & 1.44 & 0.38  \\
IRL188     &  2MASS J07464256+2000321	  &  07 46 42.56   &  +20 00 32.18    &  L0.5       &  2504$^{CTR}$ & $\ge$5.0$^{ST }$  &  0.00$^{ST }$  & 0.77 & 1.28 & 0.51  \\
\hspace{0.2cm}IRL137     &  HD 063302 ext		  &  07 47 38.52   &  $-$15 59 26.48  &  K1Ia       &  4500$^{C07}$ & 0.2 $^{C07}$  &  0.12$^{C07}$  & 0.51 & 0.60 & 0.10  \\
IRL138     &  HD 063302 		  &  07 47 38.52   &  $-$15 59 26.48  &  K1Ia       &  4500$^{C07}$ & 0.2 $^{C07}$  &  0.12$^{C07}$  & 0.67 & 0.86 & 0.19  \\
IRL288     &  HD 064332 		  &  07 53 05.27   &  $-$11 37 29.35  &  S4.5       &  3727$^{CTR}$ & 0.27$^{SEC}$  & -0.41$^{SEC}$  & 0.90 & 1.16 & 0.26  \\
IRL255     &  Gl 299			  &  08 11 57.57   &  +08 46 22.05    &  M4V        &  4055$^{CTR}$ & 4.75$^{ST }$  &  0.00$^{ST }$  & 0.48 & 0.77 & 0.29  \\
\hspace{0.2cm}IRL254     &  Gl 299 ext		  &  08 11 57.57   &  +08 46 22.05    &  M4V        &  3828$^{CTR}$ & 4.75$^{ST }$  &  0.00$^{ST }$  & 0.45 & 0.71 & 0.27  \\
IRL265     &  HD 069243 		  &  08 16 33.82   &  +11 43 34.45    &  M6III      &  3377$^{CTR}$ & 1.0 $^{ST }$  &  0.00$^{ST }$  & 0.91 & 1.46 & 0.55  \\
IRL002     &  HD 070138 		  &  08 19 43.09   &  $-$18 15 52.84  &  CJ4.5IIIa  &  3216$^{CTR}$ & -0.5$^{ST }$  &  0.00$^{ST }$  & 0.97 & 1.43 & 0.46  \\
IRL199     &  2MASS J08251968+2115521	  &  08 25 19.68   &  +21 15 52.12    &  L7.5       &  1355$^{CTR}$ & $\ge$5.0$^{ST }$  &  0.00$^{ST }$  & 1.23 & 2.02 & 0.80  \\
IRL264     &  GJ 1111			  &  08 29 49.34   &  +26 46 33.74    &  M6.5V      &  3285$^{CTR}$ & 4.85$^{ST }$  &  0.00$^{ST }$  & 0.60 & 1.00 & 0.40  \\
IRL062     &  HD 074395 		  &  08 43 40.37   &  $-$07 14 01.43  &  G1Ib       &  5250$^{C07}$ & 1.3 $^{C07}$  & -0.05$^{C07}$  & 0.40 & 0.47 & 0.07  \\
IRL031     &  HD 075555 		  &  08 52 21.79   &  +44 53 51.46    &  F5.5III    &  6745$^{FSF}$ & 3.19$^{FSF}$  & -0.22$^{FSF}$  & 0.24 & 0.28 & 0.04  \\
\hspace{0.2cm}IRL116     &  HD 075732 ext		  &  08 52 35.81   &  +28 19 50.95    &  G8V        &  5079$^{C07}$ & 4.48$^{C07}$  &  0.16$^{C07}$  & 0.37 & 0.44 & 0.07  \\
IRL117     &  HD 075732 		  &  08 52 35.81   &  +28 19 50.95    &  G8V        &  5079$^{C07}$ & 4.48$^{C07}$  &  0.16$^{C07}$  & 0.41 & 0.50 & 0.09  \\
IRL284     &  LH S2065  		  &  08 53 36.19   &  $-$03 29 32.11  &  M9V        &  2638$^{CTR}$ & 5.0 $^{ST }$  &  0.00$^{ST }$  & 0.79 & 1.32 & 0.53  \\
IRL075     &  HD 076151 		  &  08 54 17.94   &  $-$05 26 04.05  &  G2V        &  5940$^{FSF}$ & 4.23$^{FSF}$  &  0.01$^{FSF}$  & 0.34 & 0.41 & 0.08  \\
IRL005     &  HD 076221 		  &  08 55 22.88   &  +17 13 52.58    &  CN4.5      &  2782$^{CTR}$ & -0.5$^{ST }$  & -0.37$^{SEC}$  & 1.11 & 1.72 & 0.61  \\
IRL008     &  HD 076846 		  &  08 59 48.94   &  +33 46 26.47    &  CR2IIIa    &  5582$^{ ST}$ & 2.09$^{FSF}$  & -0.15$^{FSF}$  & 0.56 & 0.72 & 0.15  \\
IRL030     &  HD 087822 		  &  10 08 15.88   &  +31 36 14.58    &  F4V        &  6377$^{FSF}$ & 3.75$^{FSF}$  & -0.08$^{FSF}$  & 0.21 & 0.24 & 0.03  \\
IRL221     &  Gl 381			  &  10 12 04.69   &  $-$02 41 05.07  &  M2.5V      &  3986$^{CTR}$ & 4.7 $^{ST }$  &  0.00$^{ST }$  & 0.62 & 0.84 & 0.22  \\
IRL080     &  HD 088639 		  &  10 13 49.70   &  +27 08 08.95    &  G3IIIb     &  5431$^{FSF}$ & 2.85$^{FSF}$  & -0.41$^{FSF}$  & 0.44 & 0.50 & 0.06  \\
IRL015     &  HD 089025 		  &  10 16 41.41   &  +23 25 02.32    &  F0IIIa     &  7083$^{C07}$ & 3.2 $^{C07}$  & -0.03$^{C07}$  & 0.18 & 0.21 & 0.02  \\
IRL246     &  Gl 388			  &  10 19 36.27   &  +19 52 12.06    &  M3V        &  3988$^{CTR}$ & 4.74$^{ST }$  &  0.00$^{ST }$  & 0.62 & 0.84 & 0.22  \\
IRL186     &  HD 237903 		  &  10 30 25.31   &  +55 59 56.83    &  K7V        &  4070$^{C07}$ & 4.7 $^{C07}$  & -0.18$^{C07}$  & 0.66 & 0.78 & 0.12  \\
IRL139     &  HD 091810 		  &  10 37 20.55   &  +56 25 52.82    &  K1IIIb     &  4525$^{FSF}$ & 2.79$^{ST }$  & -0.06$^{FSF}$  & 0.59 & 0.71 & 0.12  \\
IRL004     &  HD 092055 		  &  10 37 33.27   &  $-$13 23 04.35  &  CN4.5      &  3039$^{CTR}$ & -0.5$^{C07}$  & -0.10$^{C07}$  & 1.09 & 1.54 & 0.45  \\
IRL283     &  DENIS$-$P J104814.7$-$395606 & 10 48 14.64   &  $-$39 56 06.24  &  M9V        &  3069$^{CTR}$ & 5.0 $^{ST }$  &  0.00$^{ST }$  & 0.61 & 1.05 & 0.44  \\
IRL085     &  HD 094481 		  &  10 54 17.77   &  $-$13 45 28.94  &  G4III      &  5347$^{FSF}$ & 2.87$^{FSF}$  & -0.37$^{FSF}$  & 0.45 & 0.52 & 0.07  \\
IRL256     &  HD 094705 		  &  10 56 01.46   &  +06 11 07.32    &  M5.5III    &  3561$^{CTR}$ & 0.2 $^{C07}$  & -2.50$^{C07}$  & 0.91 & 1.21 & 0.31  \\
IRL268     &  Gl 406			  &  10 56 28.86   &  +07 00 52.77    &  M6V        &  3158$^{CTR}$ & 4.84$^{ST }$  &  0.00$^{ST }$  & 0.64 & 1.03 & 0.39  \\
IRL066     &  HD 095128 		  &  10 59 27.97   &  +40 25 48.92    &  G1V        &  5834$^{C07}$ & 4.34$^{C07}$  &  0.04$^{C07}$  & 0.31 & 0.35 & 0.04  \\
IRL231     &  HD 095735 		  &  11 03 20.19   &  +35 58 11.56    &  M2V        &  4050$^{CTR}$ & 4.8 $^{C07}$  & -0.20$^{C07}$  & 0.57 & 0.77 & 0.20  \\
\hspace{0.2cm}IRL157     &  HD 099998 ext		  &  11 30 18.89   &  $-$03 00 12.59  &  K3III      &  4176$^{CTR}$ & 1.67$^{C07}$  & -0.39$^{C07}$  & 0.73 & 0.86 & 0.13  \\
IRL158     &  HD 099998 		  &  11 30 18.89   &  $-$03 00 12.59  &  K3III      &  3930$^{CTR}$ & 1.67$^{C07}$  & -0.39$^{C07}$  & 0.82 & 0.99 & 0.17  \\
IRL131     &  HD 100006 		  &  11 30 29.03   &  +18 24 35.28    &  K0III      &  4810$^{FSF}$ & 2.93$^{ST }$  & -0.32$^{FSF}$  & 0.59 & 0.67 & 0.09  \\
IRL115     &  HD 101501 		  &  11 41 03.01   &  +34 12 05.88    &  G8V        &  5401$^{C07}$ & 4.6 $^{C07}$  & -0.13$^{C07}$  & 0.37 & 0.42 & 0.05  \\
IRL192     &  2MASS J11463449+2230527	  &  11 46 34.49   &  +22 30 52.74    &  L3         &  2170$^{CTR}$ & $\ge$5.0$^{ST }$  &  0.00$^{ST }$  & 0.95 & 1.57 & 0.62  \\
IRL044     &  HD 102870 		  &  11 50 41.71   &  +01 45 52.99    &  F8.5IV     &  6109$^{C07}$ & 4.2 $^{C07}$  &  0.17$^{C07}$  & 0.28 & 0.32 & 0.03  \\
IRL112     &  HD 104979 		  &  12 05 12.54   &  +08 43 58.74    &  G8III      &  5030$^{FSF}$ & 2.35$^{FSF}$  & -0.13$^{FSF}$  & 0.54 & 0.62 & 0.08  \\
IRL281     &  BR B1219$-$1336		  &  12 21 52.48   &  $-$13 53 10.20  &  M9III      &  3487$^{CTR}$ & 0.6 $^{ST }$  &  0.00$^{ST }$  & 0.76 & 1.28 & 0.52  \\
IRL018     &  HD 108519 		  &  12 27 46.30   &  +27 25 21.95    &  F0V        &  6608$^{FSF}$ & 4.3 $^{ST }$  &  0.00$^{ST }$  & 0.16 & 0.20 & 0.04  \\
IRL084     &  HD 108477 		  &  12 27 49.44   &  $-$16 37 54.63  &  G4III      &  5442$^{FSF}$ & 3.25$^{ST }$  & -0.05$^{FSF}$  & 0.45 & 0.53 & 0.09  \\
IRL270     &  HD 108849 		  &  12 30 21.01   &  +04 24 59.16    &  M7III      &  3491$^{CTR}$ & 0.84$^{ST }$  &  0.00$^{ST }$  & 0.92 & 1.30 & 0.38  \\
IRL060     &  HD 109358 		  &  12 33 44.54   &  +41 21 26.92    &  G0V        &  5986$^{FSF}$ & 4.11$^{FSF}$  & -0.13$^{FSF}$  & 0.33 & 0.35 & 0.02  \\
IRL050     &  HD 111844 		  &  12 51 54.38   &  +19 10 05.06    &  F8IV       &  6524$^{FSF}$ & 3.92$^{ST }$  & -0.74$^{SEC}$  & 0.13 & 0.13 & 0.00  \\
IRL291     &  SDSS J125453.90$-$012247.4  &  12 54 53.93   &  $-$01 22 47.49  &  T2         &  1449$^{CTR}$ & $\ge$5.0$^{ST }$  &  0.00$^{ST }$  & 0.76 & 0.92 & 0.16  \\
IRL026     &  HD 113139 		  &  13 00 43.69   &  +56 21 58.81    &  F2V        &  6810$^{C07}$ & 3.87$^{C07}$  &  0.22$^{C07}$  & 0.21 & 0.22 & 0.02  \\
IRL191     &  Kelu			  &  13 05 40.19   &  $-$25 41 05.99  &  L2         &  2125$^{CTR}$ & $\ge$5.0$^{ST }$  &  0.00$^{ST }$  & 0.95 & 1.60 & 0.66  \\
IRL053     &  HD 114710 		  &  13 11 52.39   &  +27 52 41.45    &  F9.5V      &  5975$^{C07}$ & 4.4 $^{C07}$  &  0.09$^{C07}$  & 0.22 & 0.21 & -0.01 \\
IRL154     &  HD 114960 		  &  13 13 57.56   &  +01 27 23.20    &  K3.5IIIb   &  4331$^{FSF}$ & 1.97$^{FSF}$  & -0.13$^{FSF}$  & 0.68 & 0.82 & 0.13  \\
IRL108     &  HD 114946 		  &  13 14 10.89   &  $-$19 55 51.40  &  G7IV       &  5171$^{C07}$ & 3.64$^{C07}$  &  0.13$^{C07}$  & 0.50 & 0.58 & 0.08  \\
IRL094     &  HD 115617 		  &  13 18 24.31   &  $-$18 18 40.30  &  G6.5V      &  5531$^{C07}$ & 4.32$^{C07}$  & -0.01$^{C07}$  & 0.32 & 0.33 & 0.01  \\
IRL287     &  BD+442267 		  &  13 21 18.73   &  +43 59 13.67    &  S2.5       &  3926$^{CTR}$ & 0.55$^{ST }$  &  0.00$^{ST }$  & 0.84 & 1.07 & 0.22  \\
IRL229     &  HD 120052 		  &  13 47 25.39   &  $-$17 51 35.42  &  M2III      &  3608$^{CTR}$ & 1.35$^{ST }$  &  0.00$^{ST }$  & 0.91 & 1.16 & 0.25  \\
IRL176     &  HD 120477 		  &  13 49 28.64   &  +15 47 52.46    &  K5.5III    &  3913$^{CTR}$ & 1.32$^{SEC}$  & -0.30$^{SEC}$  & 0.81 & 1.00 & 0.19  \\
IRL114     &  HD 122563 		  &  14 02 31.84   &  +09 41 09.94    &  G8III      &  4566$^{C07}$ & 1.12$^{C07}$  & -2.63$^{C07}$  & 0.58 & 0.68 & 0.10  \\
IRL206     &  IRAS 14086$-$0730		  &  14 11 18.03   &  $-$07 44 47.30  &  M10III     &  3601$^{CTR}$ & 0.5 $^{ST }$  &  0.00$^{ST }$  & 2.15 & 4.12 & 1.98  \\
\hspace{0.2cm}IRL135     &  HD 124897 shape ext	  &  14 15 39.67   &  +19 10 56.67    &  K1.5III    &  4361$^{C07}$ & 1.93$^{C07}$  & -0.53$^{C07}$  & 0.68 & 0.77 & 0.09  \\
IRL136     &  HD 124897 shape		  &  14 15 39.67   &  +19 10 56.67    &  K1.5III    &  4361$^{C07}$ & 1.93$^{C07}$  & -0.53$^{C07}$  & 0.71 & 0.82 & 0.11  \\
IRL134     &  HD 124897 lines		  &  14 15 39.67   &  +19 10 56.67    &  K1.5III    &  4361$^{C07}$ & 1.93$^{C07}$  & -0.53$^{C07}$  & 0.71 & 0.82 & 0.11  \\
\hspace{0.2cm}IRL133     &  HD 124897 lines ext	  &  14 15 39.67   &  +19 10 56.67    &  K1.5III    &  4361$^{C07}$ & 1.93$^{C07}$  & -0.53$^{C07}$  & 0.68 & 0.77 & 0.09  \\
IRL042     &  HD 124850 		  &  14 16 00.86   &  $-$06 00 01.96  &  F7III      &  6116$^{C07}$ & 3.83$^{C07}$  & -0.11$^{C07}$  & 0.28 & 0.30 & 0.02  \\
IRL043     &  HD 126660 		  &  14 25 11.79   &  +51 51 02.67    &  F7V        &  6202$^{C07}$ & 3.84$^{C07}$  & -0.27$^{C07}$  & 0.24 & 0.30 & 0.05  \\
IRL074     &  HD 126868 		  &  14 28 12.13   &  $-$02 13 40.65  &  G2IV       &  5731$^{FSF}$ & 3.35$^{FSF}$  & -0.15$^{FSF}$  & 0.36 & 0.42 & 0.06  \\
IRL285     &  LH S2924  		  &  14 28 43.23   &  +33 10 39.14    &  M9V        &  2772$^{CTR}$ & 5.0 $^{ST }$  &  0.00$^{ST }$  & 0.76 & 1.26 & 0.50  \\
IRL274     &  IRAS 14303$-$1042		  &  14 32 59.89   &  $-$10 56 03.60  &  M8III      &  3306$^{CTR}$ & 0.7 $^{ST }$  &  0.00$^{ST }$  & 0.79 & 1.38 & 0.59  \\
IRL190     &  2MASS J14392836+1929149	  &  14 39 28.36   &  +19 29 14.98    &  L1         &  2528$^{CTR}$ & $\ge$5.0$^{ST }$  &  0.00$^{ST }$  & 0.75 & 1.25 & 0.50  \\
IRL275     &  IRAS 14436$-$0703		  &  14 46 18.41   &  $-$07 15 49.80  &  M8III      &  3465$^{CTR}$ & 0.7 $^{ST }$  &  0.00$^{ST }$  & 0.76 & 1.29 & 0.53  \\
IRL149     &  HD 132935 		  &  15 02 04.23   &  $-$08 20 40.99  &  K2III      &  4547$^{FSF}$ & 2.17$^{FSF}$  & -0.11$^{FSF}$  & 0.73 & 0.87 & 0.15  \\
\hspace{0.2cm}IRL148     &  HD 132935 ext		  &  15 02 04.23   &  $-$08 20 40.99  &  K2III      &  4502$^{FSF}$ & 2.68$^{ST }$  & -0.07$^{FSF}$  & 0.66 & 0.78 & 0.11  \\
IRL193     &  2MASS J15065441+1321060	  &  15 06 54.41   &  +13 21 06.08    &  L3         &  2064$^{CTR}$ & $\ge$5.0$^{ST }$  &  0.00$^{ST }$  & 0.97 & 1.65 & 0.67  \\
IRL196     &  2MASS J15074769$-$1627386	  &  15 07 47.69   &  $-$16 27 38.62  &  L5         &  2234$^{CTR}$ & $\ge$5.0$^{ST }$  &  0.00$^{ST }$  & 0.96 & 1.53 & 0.57  \\
IRL282     &  IRAS 15060+0947		  &  15 08 25.77   &  +09 36 18.20    &  M9III      &  2203$^{CTR}$ & 0.6 $^{ST }$  &  0.00$^{ST }$  & 1.30 & 2.45 & 1.15  \\
IRL012     &  HD 135153 		  &  15 14 37.31   &  $-$31 31 08.85  &  F0Ib       &  7073$^{FSF}$ & 2.07$^{FSF}$  & -0.06$^{SEC}$  & 0.21 & 0.29 & 0.08  \\
\hspace{0.2cm}IRL011     &  HD 135153 ext		  &  15 14 37.31   &  $-$31 31 08.85  &  F0Ib       &  7037$^{FSF}$ & 2.03$^{FSF}$  & -0.06$^{SEC}$  & 0.15 & 0.19 & 0.04  \\
IRL198     &  2MASS J15150083+4847416	  &  15 15 00.83   &  +48 47 41.69    &  L6         &  2014$^{CTR}$ & $\ge$5.0$^{ST }$  &  0.00$^{ST }$  & 1.04 & 1.68 & 0.64  \\
IRL113     &  HD 135722 		  &  15 15 30.16   &  +33 18 53.39    &  G8III      &  4847$^{C07}$ & 2.56$^{C07}$  & -0.44$^{C07}$  & 0.54 & 0.65 & 0.11  \\
IRL222     &  Gl 581			  &  15 19 27.50   &  $-$07 43 19.44  &  M2.5V      &  3966$^{CTR}$ & 4.7 $^{ST }$  &  0.00$^{ST }$  & 0.59 & 0.84 & 0.26  \\
IRL150     &  HD 137759 		  &  15 24 55.77   &  +58 57 57.83    &  K2III      &  4498$^{C07}$ & 2.38$^{C07}$  &  0.05$^{C07}$  & 0.63 & 0.75 & 0.12  \\
IRL263     &  HD 142143 		  &  15 50 46.62   &  +48 28 58.85    &  M6.5III    &  3443$^{CTR}$ & 0.88$^{ST }$  &  0.00$^{ST }$  & 0.96 & 1.36 & 0.40  \\
IRL142     &  HD 142091 		  &  15 51 13.93   &  +35 39 26.56    &  K1IVa      &  4796$^{C07}$ & 3.22$^{C07}$  &  0.00$^{C07}$  & 0.51 & 0.58 & 0.07  \\
IRL132     &  HD 145675 		  &  16 10 24.31   &  +43 49 03.52    &  K0V        &  5264$^{C07}$ & 4.66$^{C07}$  &  0.34$^{C07}$  & 0.39 & 0.43 & 0.05  \\
IRL022     &  BD+382803 		  &  16 35 57.28   &  +37 58 02.10    &  F2Ib       &  6588$^{FSF}$ & 1.87$^{FSF}$  &  0.00$^{ST }$  & 0.34 & 0.41 & 0.08  \\
IRL273     &  Gl 644			  &  16 55 28.75   &  $-$08 20 10.78  &  M7V        &  3278$^{CTR}$ & 4.9 $^{ST }$  &  0.00$^{ST }$  & 0.59 & 1.00 & 0.41  \\
IRL258     &  HD 156014 		  &  17 14 38.87   &  +14 23 25.01    &  M5Ib       &  4135$^{CTR}$ & 0.76$^{C07}$  & -2.50$^{C07}$  & 0.68 & 0.76 & 0.08  \\
IRL038     &  HD 160365 		  &  17 38 57.85   &  +13 19 45.34    &  F6III      &  6070$^{C07}$ & 3.0 $^{C07}$  & -0.26$^{C07}$  & 0.20 & 0.16 & -0.04 \\
\hspace{0.2cm}IRL095     &  HD 161664 ext		  &  17 47 45.60   &  $-$22 28 40.05  &  G6Ib       &  4739$^{FSF}$ & 1.64$^{FSF}$  & -0.07$^{FSF}$  & 0.50 & 0.58 & 0.08  \\
IRL096     &  HD 161664 		  &  17 47 45.60   &  $-$22 28 40.05  &  G6Ib       &  4804$^{FSF}$ & 1.63$^{FSF}$  & -0.09$^{FSF}$  & 0.63 & 0.80 & 0.16  \\
IRL187     &  HD 164136 		  &  17 58 30.14   &  +30 11 21.38    &  F2IV       &  6799$^{C07}$ & 2.63$^{C07}$  & -0.30$^{C07}$  & 0.24 & 0.26 & 0.02  \\
\hspace{0.2cm}IRL121     &  HD 164349 ext		  &  18 00 03.41   &  +16 45 03.30    &  K0.5IIb    &  4446$^{C07}$ & 1.5 $^{C07}$  &  0.39$^{C07}$  & 0.49 & 0.52 & 0.03  \\
IRL122     &  HD 164349 		  &  18 00 03.41   &  +16 45 03.30    &  K0.5IIb    &  4446$^{C07}$ & 1.5 $^{C07}$  &  0.39$^{C07}$  & 0.53 & 0.59 & 0.06  \\
IRL143     &  HD 165438 		  &  18 06 15.19   &  $-$04 45 04.51  &  K1IV       &  4862$^{C07}$ & 3.4 $^{C07}$  &  0.02$^{C07}$  & 0.51 & 0.59 & 0.08  \\
IRL093     &  HD 165185 		  &  18 06 23.71   &  $-$36 01 11.23  &  G5V        &  5974$^{FSF}$ & 3.85$^{FSF}$  & -0.15$^{FSF}$  & 0.29 & 0.34 & 0.05  \\
IRL057     &  HD 165908 		  &  18 07 01.53   &  +30 33 43.68    &  F9V        &  5928$^{C07}$ & 4.24$^{C07}$  & -0.53$^{C07}$  & 0.26 & 0.31 & 0.05  \\
\hspace{0.2cm}IRL125     &  HD 165782 ext		  &  18 08 26.51   &  $-$18 33 07.92  &  K0Ia       &  5211$^{FSF}$ & 0.25$^{ST }$  &  0.20$^{FSF}$  & 0.34 & 0.44 & 0.10  \\
IRL126     &  HD 165782 		  &  18 08 26.51   &  $-$18 33 07.92  &  K0Ia       &  5600$^{FSF}$ & 0.25$^{ST }$  &  0.21$^{FSF}$  & 0.63 & 0.90 & 0.27  \\
IRL119     &  HD 170820 		  &  18 32 13.11   &  $-$19 07 26.28  &  G9II       &  4604$^{C07}$ & 1.62$^{C07}$  &  0.17$^{C07}$  & 0.69 & 0.89 & 0.20  \\
\hspace{0.2cm}IRL118     &  HD 170820 ext		  &  18 32 13.11   &  $-$19 07 26.28  &  G9II       &  4604$^{C07}$ & 1.62$^{C07}$  &  0.17$^{C07}$  & 0.53 & 0.63 & 0.10  \\
\hspace{0.2cm}IRL019     &  HD 173638 ext		  &  18 46 43.32   &  $-$10 07 30.17  &  F1II       &  7005$^{FSF}$ & 2.1 $^{FSF}$  & -0.68$^{FSF}$  & 0.14 & 0.14 & 0.00  \\
IRL020     &  HD 173638 		  &  18 46 43.32   &  $-$10 07 30.17  &  F1II       &  7146$^{FSF}$ & 2.2 $^{FSF}$  & -0.56$^{FSF}$  & 0.25 & 0.32 & 0.07  \\
IRL259     &  HD 175865 		  &  18 55 20.10   &  +43 56 45.93    &  M5III      &  3690$^{CTR}$ & 0.5 $^{C07}$  &  0.14$^{C07}$  & 0.87 & 1.08 & 0.21  \\
IRL056     &  HD 176051 		  &  18 57 01.60   &  +32 54 04.57    &  F9V        &  5895$^{FSF}$ & 3.91$^{FSF}$  & -0.21$^{FSF}$  & 0.36 & 0.40 & 0.05  \\
\hspace{0.2cm}IRL078     &  HD 176123 ext		  &  18 59 26.77   &  $-$18 33 59.16  &  G3II       &  5449$^{FSF}$ & 2.8 $^{FSF}$  & -0.23$^{FSF}$  & 0.42 & 0.48 & 0.06  \\
IRL079     &  HD 176123 		  &  18 59 26.77   &  $-$18 33 59.16  &  G3II       &  5302$^{FSF}$ & 2.61$^{FSF}$  & -0.30$^{FSF}$  & 0.46 & 0.54 & 0.08  \\
\hspace{0.2cm}IRL161     &  HD 178208 ext		  &  19 05 09.83   &  +49 55 23.39    &  K3III      &  4882$^{FSF}$ & 2.53$^{FSF}$  & -0.26$^{FSF}$  & 0.62 & 0.72 & 0.10  \\
IRL162     &  HD 178208 		  &  19 05 09.83   &  +49 55 23.39    &  K3III      &  4884$^{FSF}$ & 2.48$^{FSF}$  & -0.23$^{FSF}$  & 0.66 & 0.78 & 0.13  \\
\hspace{0.2cm}IRL129     &  HD 179870 ext		  &  19 13 53.58   &  +09 01 59.59    &  K0II       &  4776$^{FSF}$ & 1.53$^{FSF}$  &  0.04$^{FSF}$  & 0.44 & 0.44 & 0.00  \\
IRL130     &  HD 179870 		  &  19 13 53.58   &  +09 01 59.59    &  K0II       &  4928$^{FSF}$ & 1.57$^{FSF}$  &  0.06$^{FSF}$  & 0.49 & 0.53 & 0.04  \\
\hspace{0.2cm}IRL086     &  HD 179821 ext		  &  19 13 58.60   &  +00 07 31.92    &  G4Ia       &  5708$^{CTR}$ & 1.08$^{FSF}$  &  0.00$^{ST }$  & 0.25 & 0.38 & 0.13  \\
IRL087     &  HD 179821 		  &  19 13 58.60   &  +00 07 31.92    &  G4Ia       &  4645$^{CTR}$ & 0.95$^{FSF}$  &  0.00$^{ST }$  & 0.43 & 0.67 & 0.24  \\
IRL278     &  Gl 752			  &  19 16 55.26   &  +05 10 08.10    &  M8V        &  2970$^{CTR}$ & 4.96$^{ST }$  &  0.00$^{ST }$  & 0.64 & 1.09 & 0.45  \\
IRL179     &  HD 181596 		  &  19 18 30.12   &  +50 13 39.42    &  K5III      &  4275$^{FSF}$ & 2.35$^{ST }$  & -0.41$^{FSF}$  & 0.80 & 0.98 & 0.18  \\
\hspace{0.2cm}IRL182     &  HD 181475 ext		  &  19 20 48.31   &  $-$04 30 09.01  &  K7IIa      &  3996$^{FSF}$ & 0.65$^{FSF}$  & -0.37$^{FSF}$  & 0.78 & 0.96 & 0.19  \\
IRL183     &  HD 181475 		  &  19 20 48.31   &  $-$04 30 09.01  &  K7IIa      &  4124$^{FSF}$ & 0.83$^{FSF}$  & -0.21$^{FSF}$  & 0.94 & 1.22 & 0.28  \\
IRL107     &  HD 182694 		  &  19 23 56.50   &  +43 23 17.41    &  G7III      &  5312$^{FSF}$ & 2.63$^{FSF}$  & -0.19$^{FSF}$  & 0.45 & 0.55 & 0.10  \\
IRL024     &  HD 182835 		  &  19 26 31.09   &  +00 20 18.85    &  F2Ib       &  7350$^{C07}$ & 2.15$^{C07}$  &  0.09$^{C07}$  & 0.25 & 0.31 & 0.06  \\
\hspace{0.2cm}IRL023     &  HD 182835 ext		  &  19 26 31.09   &  +00 20 18.85    &  F2Ib       &  7350$^{C07}$ & 2.15$^{C07}$  &  0.09$^{C07}$  & 0.12 & 0.11 & -0.01 \\
IRL059     &  HD 185018 		  &  19 36 52.45   &  +11 16 23.53    &  G0Ib       &  5550$^{C07}$ & 1.3 $^{C07}$  & -0.24$^{C07}$  & 0.41 & 0.48 & 0.07  \\
\hspace{0.2cm}IRL058     &  HD 185018 ext		  &  19 36 52.45   &  +11 16 23.53    &  G0Ib       &  5550$^{C07}$ & 1.3 $^{C07}$  & -0.24$^{C07}$  & 0.37 & 0.43 & 0.05  \\
\hspace{0.2cm}IRL168     &  HD 185622 ext		  &  19 39 25.33   &  +16 34 16.03    &  K4Ib       &  3736$^{FSF}$ & 1.05$^{ST }$  & -0.19$^{FSF}$  & 0.74 & 0.92 & 0.18  \\
IRL169     &  HD 185622 		  &  19 39 25.33   &  +16 34 16.03    &  K4Ib       &  3898$^{FSF}$ & 0.63$^{FSF}$  & -0.14$^{FSF}$  & 0.90 & 1.17 & 0.27  \\
IRL035     &  HD 186155 		  &  19 40 50.18   &  +45 31 29.78    &  F5II       &  6590$^{FSF}$ & 3.34$^{FSF}$  &  0.31$^{FSF}$  & 0.18 & 0.22 & 0.03  \\
IRL156     &  HD 187238 		  &  19 48 11.83   &  +22 45 46.35    &  K3Iab      &  4143$^{FSF}$ & 0.99$^{FSF}$  & -0.01$^{FSF}$  & 0.81 & 1.02 & 0.21  \\
\hspace{0.2cm}IRL155     &  HD 187238 ext		  &  19 48 11.83   &  +22 45 46.35    &  K3Iab      &  4228$^{FSF}$ & 0.77$^{FSF}$  & -0.09$^{FSF}$  & 0.63 & 0.73 & 0.10  \\
\hspace{0.2cm}IRL214     &  HD 339034 ext		  &  19 50 11.93   &  +24 55 24.19    &  M1Ia       &  3960$^{CTR}$ & -0.0$^{ST }$  &  0.00$^{ST }$  & 0.64 & 0.87 & 0.22  \\
IRL215     &  HD 339034 		  &  19 50 11.93   &  +24 55 24.19    &  M1Ia       &  3342$^{CTR}$ & -0.0$^{ST }$  &  0.00$^{ST }$  & 1.05 & 1.51 & 0.46  \\
\hspace{0.2cm}IRL089     &  HD 190113 ext		  &  20 02 02.85   &  +35 38 28.01    &  G5Ib       &  4968$^{FSF}$ & 1.45$^{FSF}$  & -0.28$^{FSF}$  & 0.45 & 0.49 & 0.05  \\
IRL090     &  HD 190113 		  &  20 02 02.85   &  +35 38 28.01    &  G5Ib       &  4996$^{FSF}$ & 1.59$^{FSF}$  & -0.31$^{FSF}$  & 0.60 & 0.73 & 0.14  \\
IRL104     &  HD 333385 		  &  20 02 27.37   &  +30 04 25.46    &  G7Ia       &  5375$^{FSF}$ & 0.36$^{ST }$  &  0.14$^{FSF}$  & 0.61 & 0.90 & 0.29  \\
\hspace{0.2cm}IRL103     &  HD 333385 ext		  &  20 02 27.37   &  +30 04 25.46    &  G7Ia       &  5475$^{FSF}$ & 1.05$^{FSF}$  &  0.21$^{FSF}$  & 0.29 & 0.38 & 0.10  \\
IRL046     &  HD 190323 		  &  20 03 49.61   &  +14 58 58.74    &  F8Ia       &  5740$^{CTR}$ & 0.81$^{ST }$  &  0.07$^{FSF}$  & 0.32 & 0.38 & 0.06  \\
\hspace{0.2cm}IRL045     &  HD 190323 ext		  &  20 03 49.61   &  +14 58 58.74    &  F8Ia       &  6906$^{FSF}$ & 0.81$^{ST }$  &  0.09$^{FSF}$  & 0.22 & 0.22 & 0.00  \\
IRL077     &  HD 192713 		  &  20 15 30.23   &  +23 30 32.05    &  G3Ib       &  4864$^{C07}$ & 0.1 $^{C07}$  & -0.43$^{C07}$  & 0.48 & 0.57 & 0.09  \\
\hspace{0.2cm}IRL076     &  HD 192713 ext		  &  20 15 30.23   &  +23 30 32.05    &  G3Ib       &  4864$^{C07}$ & 0.1 $^{C07}$  & -0.43$^{C07}$  & 0.44 & 0.50 & 0.07  \\
IRL184     &  HD 194193 		  &  20 22 45.29   &  +41 01 33.64    &  K7III      &  4025$^{FSF}$ & 1.42$^{FSF}$  & -0.14$^{FSF}$  & 0.88 & 1.09 & 0.20  \\
IRL091     &  HD 193896 		  &  20 23 00.79   &  $-$09 39 16.95  &  G5IIIa     &  5332$^{FSF}$ & 3.2 $^{ST }$  & -0.49$^{FSF}$  & 0.50 & 0.61 & 0.11  \\
\hspace{0.2cm}IRL244     &  RW Cyg ext		  &  20 28 50.59   &  +39 58 54.42    &  M3Iab      &  3844$^{CTR}$ & 0.16$^{ST }$  &  0.00$^{ST }$  & 0.74 & 0.95 & 0.21  \\
IRL245     &  RW Cyg			  &  20 28 50.59   &  +39 58 54.42    &  M3Iab      &  3341$^{CTR}$ & 0.16$^{ST }$  &  0.00$^{ST }$  & 1.09 & 1.51 & 0.42  \\
IRL267     &  HD 196610 		  &  20 37 54.72   &  +18 16 06.89    &  M6III      &  3643$^{CTR}$ & 0.91$^{ST }$  &  0.00$^{ST }$  & 0.86 & 1.12 & 0.26  \\
IRL230     &  Gl 806			  &  20 45 04.09   &  +44 29 56.66    &  M2V        &  3660$^{FSF}$ & 4.65$^{FSF}$  & -0.23$^{FSF}$  & 0.62 & 0.78 & 0.17  \\
\hspace{0.2cm}IRL170     &  HD 201065 ext		  &  21 05 35.78   &  +46 57 47.76    &  K4Ib       &  4068$^{FSF}$ & 1.16$^{FSF}$  & -0.50$^{FSF}$  & 0.73 & 0.88 & 0.15  \\
IRL171     &  HD 201065 		  &  21 05 35.78   &  +46 57 47.76    &  K4Ib       &  4226$^{FSF}$ & 0.87$^{FSF}$  & -0.47$^{FSF}$  & 0.83 & 1.04 & 0.20  \\
IRL041     &  HD 201078 		  &  21 06 30.24   &  +31 11 04.76    &  F7II       &  6157$^{C07}$ & 1.65$^{C07}$  &  0.13$^{C07}$  & 0.26 & 0.33 & 0.07  \\
IRL185     &  HD 201092 		  &  21 06 55.26   &  +38 44 31.40    &  K7V        &  4348$^{CTR}$ & 4.37$^{SEC}$  & -0.72$^{SEC}$  & 0.59 & 0.73 & 0.14  \\
IRL098     &  HD 202314 		  &  21 14 10.28   &  +29 54 03.45    &  G6Ib       &  4864$^{C07}$ & 1.3 $^{C07}$  & -0.05$^{C07}$  & 0.51 & 0.66 & 0.14  \\
\hspace{0.2cm}IRL097     &  HD 202314 ext		  &  21 14 10.28   &  +29 54 03.45    &  G6Ib       &  4864$^{C07}$ & 1.3 $^{C07}$  & -0.05$^{C07}$  & 0.48 & 0.60 & 0.12  \\
IRL247     &  HD 204585 		  &  21 28 59.77   &  +22 10 45.96    &  M4.5IIIa   &  3700$^{CTR}$ & 1.11$^{ST }$  &  0.00$^{ST }$  & 0.86 & 1.07 & 0.21  \\
IRL216     &  HD 204724 		  &  21 29 56.89   &  +23 38 19.81    &  M1III      &  3765$^{CTR}$ & 1.5 $^{ST }$  &  0.00$^{ST }$  & 0.78 & 1.01 & 0.23  \\
IRL276     &  IRAS 21284$-$0747		  &  21 31 06.51   &  $-$07 34 20.50  &  M8III      &  3130$^{CTR}$ & 0.7 $^{ST }$  &  0.00$^{ST }$  & 0.87 & 1.48 & 0.61  \\
IRL224     &  HD 206936 		  &  21 43 30.46   &  +58 46 48.16    &  M2Ia       &  3540$^{CTR}$ & 0.0 $^{ST }$  &  0.00$^{ST }$  & 0.86 & 1.24 & 0.38  \\
\hspace{0.2cm}IRL223     &  HD 206936 ext		  &  21 43 30.46   &  +58 46 48.16    &  M2Ia       &  3879$^{CTR}$ & 0.0 $^{ST }$  &  0.00$^{ST }$  & 0.66 & 0.92 & 0.26  \\
IRL271     &  HD 207076 		  &  21 46 31.84   &  $-$02 12 45.93  &  M7III      &  2750$^{C07}$ & -0.5$^{C07}$  & -2.50$^{C07}$  & 0.78 & 1.03 & 0.25  \\
\hspace{0.2cm}IRL172     &  HD 207991 ext		  &  21 51 55.38   &  +48 26 13.59    &  K4III      &  4355$^{FSF}$ & 1.98$^{FSF}$  & -0.21$^{FSF}$  & 0.79 & 0.94 & 0.15  \\
IRL173     &  HD 207991 		  &  21 51 55.38   &  +48 26 13.59    &  K4III      &  4192$^{FSF}$ & 2.48$^{ST }$  & -0.32$^{FSF}$  & 0.84 & 1.02 & 0.18  \\
IRL111     &  HD 208606 		  &  21 55 20.59   &  +61 32 30.52    &  G8Ib       &  4718$^{FSF}$ & 0.66$^{FSF}$  & -0.16$^{FSF}$  & 0.67 & 0.86 & 0.19  \\
\hspace{0.2cm}IRL110     &  HD 208606 ext		  &  21 55 20.59   &  +61 32 30.52    &  G8Ib       &  4709$^{FSF}$ & 1.26$^{FSF}$  &  0.10$^{FSF}$  & 0.54 & 0.64 & 0.10  \\
IRL203     &  HD 209290 		  &  22 02 10.27   &  +01 24 00.82    &  M0.5V      &  3810$^{CTR}$ & 4.65$^{ST }$  &  0.00$^{ST }$  & 0.68 & 0.88 & 0.20  \\
IRL152     &  HD 212466 		  &  22 23 07.01   &  +55 57 47.62    &  K2Ia       &  3749$^{CTR}$ & 0.2 $^{ST }$  &  0.18$^{FSF}$  & 0.73 & 1.10 & 0.37  \\
\hspace{0.2cm}IRL151     &  HD 212466 ext		  &  22 23 07.01   &  +55 57 47.62    &  K2Ia       &  5018$^{FSF}$ & 0.2 $^{ST }$  &  0.02$^{FSF}$  & 0.42 & 0.61 & 0.19  \\
IRL195     &  2MASS J22244381$-$0158521	  &  22 24 43.81   &  $-$01 58 52.14  &  L4.5       &  1267$^{CTR}$ & $\ge$5.0$^{ST }$  &  0.00$^{ST }$  & 1.26 & 2.06 & 0.81  \\
IRL033     &  HD 213306 		  &  22 29 10.26   &  +58 24 54.71    &  F5Ib       &  6052$^{FSF}$ & 2.29$^{FSF}$  & -0.16$^{FSF}$  & 0.40 & 0.47 & 0.07  \\
\hspace{0.2cm}IRL032     &  HD 213306 ext		  &  22 29 10.26   &  +58 24 54.71    &  F5Ib       &  6052$^{FSF}$ & 2.27$^{FSF}$  & -0.14$^{FSF}$  & 0.34 & 0.37 & 0.03  \\
IRL021     &  HD 213135 		  &  22 29 46.02   &  $-$27 06 26.22  &  F1V        &  6404$^{FSF}$ & 4.3 $^{ST }$  & -0.54$^{FSF}$  & 0.20 & 0.21 & 0.01  \\
IRL204     &  HD 213893 		  &  22 34 35.93   &  +00 35 42.63    &  M0IIIb     &  4044$^{C07}$ & 1.6 $^{C07}$  & -0.08$^{C07}$  & 0.83 & 1.01 & 0.18  \\
\hspace{0.2cm}IRL261     &  Gl 866 ext		  &  22 38 33.72   &  $-$15 17 57.33  &  M5V        &  3728$^{CTR}$ & 4.8 $^{ST }$  &  0.00$^{ST }$  & 0.59 & 0.90 & 0.31  \\
IRL262     &  Gl 866			  &  22 38 33.72   &  $-$15 17 57.33  &  M5V        &  3369$^{CTR}$ & 4.8 $^{ST }$  &  0.00$^{ST }$  & 0.64 & 0.98 & 0.34  \\
IRL250     &  HD 214665 		  &  22 38 37.92   &  +56 47 44.28    &  M4III      &  3452$^{CTR}$ & 1.07$^{ST }$  &  0.00$^{ST }$  & 1.04 & 1.35 & 0.31  \\
IRL088     &  HD 214850 		  &  22 40 52.68   &  +14 32 56.97    &  G4V        &  5451$^{FSF}$ & 4.48$^{ST }$  & -0.34$^{FSF}$  & 0.44 & 0.53 & 0.09  \\
IRL040     &  HD 215648 		  &  22 46 41.58   &  +12 10 22.38    &  F6V        &  6167$^{C07}$ & 4.04$^{C07}$  & -0.32$^{C07}$  & 0.26 & 0.29 & 0.03  \\
IRL065     &  HD 216219 		  &  22 50 52.15   &  +18 00 07.56    &  G1II       &  5727$^{C07}$ & 3.36$^{C07}$  & -0.39$^{C07}$  & 0.30 & 0.35 & 0.05  \\
IRL272     &  MY Cep			  &  22 54 31.71   &  +60 49 38.89    &  M7I        &  2595$^{CTR}$ & -0.2$^{ST }$  &  0.00$^{ST }$  & 1.47 & 2.24 & 0.77  \\
IRL178     &  HD 216946 		  &  22 56 25.99   &  +49 44 00.75    &  K5Ib       &  3839$^{FSF}$ & 0.49$^{FSF}$  & -0.10$^{SEC}$  & 0.95 & 1.24 & 0.29  \\
\hspace{0.2cm}IRL177     &  HD 216946 ext		  &  22 56 25.99   &  +49 44 00.75    &  K5Ib       &  3817$^{FSF}$ & 0.68$^{FSF}$  & -0.10$^{SEC}$  & 0.90 & 1.16 & 0.26  \\
IRL036     &  HD 218804 		  &  23 10 27.20   &  +43 32 39.15    &  F5V        &  6261$^{C07}$ & 4.05$^{C07}$  & -0.23$^{C07}$  & 0.27 & 0.32 & 0.05  \\
IRL167     &  HD 219134 		  &  23 13 16.97   &  +57 10 06.08    &  K3V        &  4717$^{C07}$ & 4.5 $^{C07}$  &  0.05$^{C07}$  & 0.56 & 0.67 & 0.10  \\
IRL073     &  HD 219477 		  &  23 15 46.29   &  +28 14 52.43    &  G2II       &  5989$^{FSF}$ & 2.87$^{FSF}$  & -0.23$^{FSF}$  & 0.36 & 0.41 & 0.05  \\
IRL051     &  HD 219623 		  &  23 16 42.30   &  +53 12 48.51    &  F8V        &  6155$^{C07}$ & 4.17$^{C07}$  & -0.04$^{C07}$  & 0.29 & 0.34 & 0.05  \\
IRL220     &  HD 219734 		  &  23 17 44.64   &  +49 00 55.08    &  M2.5III    &  3658$^{CTR}$ & 0.9 $^{C07}$  &  0.27$^{C07}$  & 0.89 & 1.11 & 0.22  \\
IRL049     &  HD 220657 		  &  23 25 22.78   &  +23 24 14.76    &  F8III      &  6380$^{FSF}$ & 2.96$^{FSF}$  & -0.66$^{FSF}$  & 0.36 & 0.47 & 0.10  \\
IRL164     &  HD 221246 		  &  23 30 07.41   &  +49 07 59.31    &  K3III      &  4359$^{FSF}$ & 2.58$^{ST }$  & -0.35$^{FSF}$  & 0.73 & 0.85 & 0.13  \\
\hspace{0.2cm}IRL163     &  HD 221246 ext		  &  23 30 07.41   &  +49 07 59.31    &  K3III      &  4308$^{FSF}$ & 1.98$^{FSF}$  & -0.19$^{FSF}$  & 0.67 & 0.76 & 0.09  \\
IRL120     &  HD 222093 		  &  23 37 39.55   &  $-$13 03 36.86  &  G9III      &  4750$^{FSF}$ & 2.01$^{FSF}$  &  0.03$^{FSF}$  & 0.56 & 0.64 & 0.09  \\
IRL269     &  BRI B2339$-$0447		  &  23 42 02.75   &  $-$04 31 04.88  &  M7III      &  3425$^{CTR}$ & 0.84$^{ST }$  &  0.00$^{ST }$  & 0.93 & 1.39 & 0.46  \\
\end{longtable}
\end{footnotesize}

\begin{table}
	\caption{Additional stars from the MILES and CaT stellar libraries used as templates in the determination of the IRTF stellar temperatures, gravities and metallicities. Their stellar parameters were determined by \citet{cenarro_et_al_2007}.}
	\label{additional_templates}
	\begin{tabular}{ccrr}
		\hline
		\hline
		{\bf Star} & {\bf $T_{\mathrm{eff}}$ (K)} & {\bf $\log g$} & {\bf $[Z/Z_{\odot}]$ } \\ 
		\hline
		\hline
		BD442051 & 3696 & 5.00 & $-$1.50 \\ 
		HD058521 & 3238 & 0.00 & $-$0.19 \\ 
		HD069267 & 4043 & 1.51 & $-$0.12 \\ 
		HD073394 & 4500 & 1.10 & $-$1.38 \\ 
		HD073593 & 4717 & 2.25 & $-$0.12 \\ 
		HD076813 & 6072 & 4.20 & $-$0.82 \\ 
		HD078712 & 3202 & 0.00 & $-$0.11 \\ 
		HD079452 & 4829 & 2.35 & $-$0.84 \\ 
		HD081192 & 4705 & 2.50 & $-$0.62 \\ 
		HD083425 & 4120 & 2.00 & $-$0.35 \\ 
		HD083618 & 4231 & 1.74 & $-$0.08 \\ 
		HD083632 & 4214 & 1.00 & $-$1.39 \\ 
		HD087737 & 9625 & 1.98 & $-$0.04 \\ 
		HD095735 & 3551 & 4.90 & $-$0.20 \\ 
		HD096360 & 3550 & 0.50 & $-$0.58 \\ 
		HD099998 & 3863 & 1.79 & $-$0.16 \\ 
		HD103095 & 5025 & 4.56 & $-$1.36 \\ 
		HD107213 & 6298 & 4.01 & 0.36 \\ 
		HD111631 & 3785 & 4.75 & 0.10 \\ 
		HD114038 & 4530 & 2.71 & $-$0.04 \\ 
		HD114961 & 3012 & 0.00 & $-$0.81 \\ 
		HD119228 & 3600 & 1.60 & 0.30 \\ 
		HD119667 & 3700 & 1.00 & $-$0.35 \\ 
		HD120933 & 3820 & 1.52 & 0.50 \\ 
		HD121299 & 4710 & 2.64 & $-$0.03 \\ 
		HD123657 & 3450 & 0.85 & 0.00 \\ 
		HD126327 & 2819 & 0.00 & $-$0.58 \\ 
		HD130705 & 4336 & 2.10 & 0.41 \\ 
		HD131430 & 4190 & 2.18 & 0.04 \\ 
		HD134063 & 4885 & 2.34 & $-$0.69 \\ 
		HD136726 & 4120 & 2.03 & 0.07 \\ 
		HD137704 & 4095 & 1.97 & $-$0.27 \\ 
		HD138481 & 3890 & 1.64 & 0.20 \\ 
		HD145675 & 5264 & 4.66 & 0.34 \\ 
		HD147923 & 3600 & 0.80 & $-$0.19 \\ 
		HD148783 & 3279 & 0.20 & $-$0.06 \\ 
		HD149661 & 5168 & 4.63 & 0.04 \\ 
		HD154733 & 4279 & 2.10 & 0.00 \\ 
		HD164058 & 3930 & 1.26 & $-$0.05 \\ 
		HD167768 & 5235 & 1.61 & $-$0.68 \\ 
		HD168720 & 3810 & 1.10 & 0.00 \\ 
		HD184499 & 5738 & 4.02 & $-$0.66 \\ 
		HD184786 & 3467 & 0.60 & $-$0.04 \\ 
		HD185144 & 5260 & 4.55 & $-$0.24 \\ 
		HD187216 & 3500 & 0.40 & $-$2.48 \\ 
		HD191277 & 4459 & 2.71 & 0.30 \\ 
		HD199799 & 3400 & 0.30 & $-$0.24 \\ 
		HD232078 & 4008 & 0.30 & $-$1.73 \\
		\hline
		\hline
	\end{tabular}
\end{table}


\end{document}